\documentclass[%
notitlepage,
showpacs, preprintnumbers,
amsmath, amssymb,
aps, longbibliography,
]{revtex4-1}

\usepackage{graphicx}
\usepackage{epstopdf}
\usepackage{dcolumn}
\usepackage{bm}

\usepackage{psfrag}
\usepackage{natbib}
\usepackage{tikz}
\usepackage{upgreek}
\usepackage{multirow}
\usepackage{subfigure}
\usepackage{color}
\usepackage{amsmath}
\usepackage{amssymb}
\usepackage{afterpage}
\usepackage{array}
\usepackage{floatrow}
\newfloatcommand{capbtabbox}{table}[][\FBwidth]
\usepackage[colorlinks=true, linkcolor=blue, urlcolor=blue, citecolor=blue]{hyperref}

\definecolor{blue2}{rgb}{0, 0.4, 0.7}
\definecolor{green2}{rgb}{0, 0.6, 0}
\begin{document}

\preprint{PHYSICAL REVIEW FLUIDS}

\title{A spectral-scaling based extension to the attached eddy model of wall-turbulence} 

\author{Dileep Chandran}
	\email{dileep.padinjare@unimelb.edu.au}
	\affiliation{Department of Mechanical Engineering, University of Melbourne, Victoria, 3010, Australia.}
\author{Jason P. Monty}
	\affiliation{Department of Mechanical Engineering, University of Melbourne, Victoria, 3010, Australia.}
\author{Ivan Marusic}
	\affiliation{Department of Mechanical Engineering, University of Melbourne, Victoria, 3010, Australia.}


\begin{abstract}
Two-dimensional (2-D) spectra of the streamwise velocity component, measured at friction Reynolds numbers ranging from 2400 to 26000, are used to refine a model for the logarithmic region of turbulent boundary layers. Here, we focus on the attached eddy model (AEM).
The conventional AEM assumes the boundary layer to be populated with hierarchies of self-similar wall-attached ($Type\,A$) eddies alone. While $Type\,A$ eddies represent the dominant energetic large-scale motions at high Reynolds numbers, the scales that are not represented by such eddies are observed to carry a significant proportion of the total kinetic energy. Therefore, in the present study, we propose an extended AEM that incorporates two additional representative eddies. These eddies, named $Type\,C_A$ and $Type\,SS$, represent the self-similar but wall-incoherent low-Reynolds number features, and the non-self-similar wall-coherent superstructures, respectively. The extended AEM is shown to better predict a greater range of energetic length scales and capture the low- and high-Reynolds number scaling trends in the 2-D spectra of all three velocity components. A discussion on spectral self-similarity and the associated $k^{-1}$ scaling law is also presented. 
\end{abstract}

\maketitle
 
\section{Introduction} \label{intro}
The logarithmic region, or the inertial sublayer, is the most important region within a turbulent boundary layer at high Reynolds number, owing to its significant contribution to the overall production of turbulent kinetic energy  \citep{marusic2010highRe,jimenez2012cascades}. The importance of this inertia-dominated region has motivated several studies to characterize the coherent energy-containing motions, or `eddies', that reside within this region of the boundary layer \citep{robinson1991coherent,adrian2007hairpin,hutchins2007evidence,smits2011high,jimenez2012cascades,lozano2014time,jimenez2018coherent}.
It is the notion of self-similarity of such eddies that underpins a number of models for the logarithmic region of wall-bounded turbulent flows. Among such models, those that are based on the attached eddy hypothesis of Townsend \citep{townsend1980structure} have gained significant popularity (see references  \citealp{vassilicos2015streamwise,hwang2015statistical,yang2017structure,agostini2017spectral,mouri2017two,hwang2018wall,cho2018scale,lozano-duran_bae_2019,mckeon2019self,yoon2020wall,hwang2020attached}, among others). 

The attached eddy hypothesis assumes the boundary layer as a random distribution of ``persistent, organized flow patterns", that are influenced by the wall and whose size scale with distance from the wall. Based on the hypothesis, \citet{perry1982mechanism} developed an attached eddy model (AEM) by prescribing physical shapes to the self-similar structures. The key feature of the AEM is the concept of a `representative attached eddy' and the boundary layer is modeled as an assemblage of discrete but self-similar hierarchies of such eddies. Following Townsend's hypothesis \citep{townsend1980structure}, the size and population density of the eddies are directly and inversely proportional to their distance from the wall, respectively. Further, based on dimensional analysis, \citet{perry1982mechanism} reported such hierarchies of geometrically self-similar eddies, over a range of length scales, to contribute equally to the pre-multiplied turbulent kinetic energy. The authors hence proposed a $k_x^{-1}$ scaling in the one-dimensional (1-D) energy spectra as a characteristic of self-similarity. Here $k_x$ is the streamwise wavenumber. A refinement to the Perry and Chong model was made by \citet{perry1986theoretical} where the discrete hierarchical organization was replaced with a continuous distribution of eddies, whose sizes varied from 100 viscous units to the order of boundary layer thickness. Following that, several refinements were made to the AEM based on experimental observations, in order to better predict the Reynolds stresses, energy spectra, structure functions and higher-order moments (eg. \citealp{marusic1995wall,marusic2001role,woodcock2015statistical,de2016influence,baidya2017distance,yang2017structure,eich2020towardsAEM}). A comprehensive review of the AEM and the various refinements made to the model to date is provided by \citet{marusic2019attached}.

Insights on the three dimensional geometry of self-similar eddies in turbulent boundary layers have been obtained from recent high-Reynolds number multi-point measurements. \citet{baars2017JFM} studied spectral coherence of synchronous near-wall and outer-region velocity signals and reported that the structures that are coherent with the wall have a streamwise/wall-normal aspect ratio of $\lambda_x/z \approx 14$. Here, $\lambda_x$ and $z$ denote streamwise wavelength and wall-normal distance respectively. \citet{chandran2017JFM} conducted two-point measurements of streamwise velocity ($u$) in the streamwise/spanwise plane in order to compute the 2-D spectra as a function of streamwise and spanwise wavelengths, $\lambda_x$ and $\lambda_y$ respectively. At high Reynolds numbers, they observed that 
the smaller length scales retained the low Reynolds number behavior with a nominal $\lambda_y/z \sim (\lambda_x/z)^{1/2}$ relationship between the streamwise and spanwise length scales. In contrast, the larger length scales ($\lambda_x,\lambda_y > \mathcal{O}(10z)$) were observed to transition towards a $\lambda_y/z \sim \lambda_x/z$ scaling that is representative of self-similarity. They reported that the self-similar large-scales have a streamwise/spanwise aspect ratio of $\lambda_x/\lambda_y \approx 7$ and their streamwise/wall-normal aspect ratio agreed with that of the wall-coherent motions of \citet{baars2017JFM}.  More recently, \citet{baidya2019simultaneous} extended the model of \citet{baars2017JFM} by also capturing the azimuthal/spanwise information of wall-coherent structures in pipe and boundary layer flows at high Reynolds numbers and reported the self-similar structures to follow an aspect ratio of 7 : 1 : 1 in the streamwise, spanwise and wall-normal directions, respectively. 

\begin{figure}
	\centering
	\subfigure{		
		\includegraphics[trim = 0mm 0mm 0mm 0mm, clip, width=13cm]{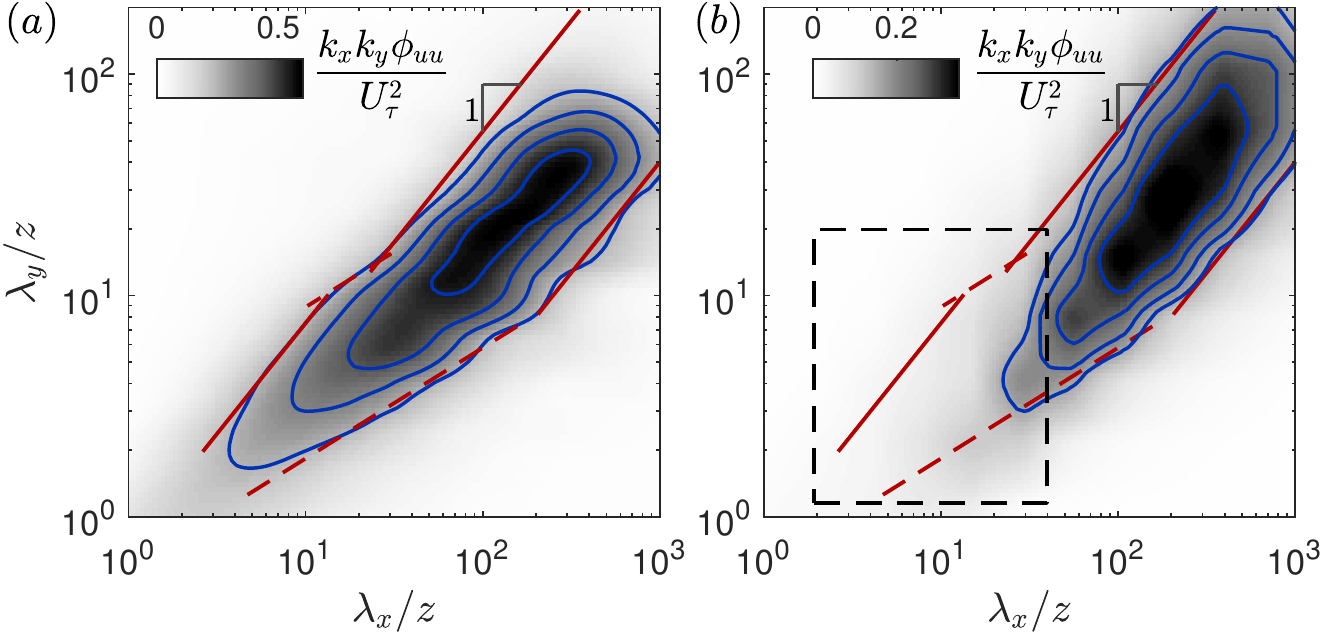}}
	\caption{(a) 2-D spectrum of the streamwise velocity ($u$) at  $z^+=2.6 {Re_\tau}^{1/2}$ for  $Re_\tau \approx 26000$; the blue contour lines represent $k_xk_y\phi_{uu}/U_\tau^2 = 0.25,\, 0.35$ and $0.45$. (b) 2-D spectrum obtained from attached eddy model with $Type\,A$ eddies at same $z^+$ and $Re_\tau$. The red solid and dashed lines denote the $\lambda_y/z \sim \lambda_x/z$ scaling and $\lambda_y/z \sim (\lambda_x/z)^{1/2}$ relationship, respectively.}
	\label{lowhigh}
\end{figure}

Adopting the empirically observed aspect ratio of self-similar eddies, \citet{chandran2017JFM} showed that hierarchies of self-similar wall-attached eddies, named $Type\,A$ (following the notation of \citet{marusic1995wall}), represented the energetic large-scale region of the 2-D spectrum reasonably well. This result is reviewed in figure \ref{lowhigh}, where the 2-D spectrum obtained from the AEM is compared against the experimental data at a friction Reynolds number of $Re_\tau = 26000$. (Here, $Re_\tau = \delta U_\tau/\nu$, where $\delta$ is the boundary layer thickness, $U_\tau$ is the friction velocity and $\nu$ is the kinematic viscosity.)
The model, however, does not capture the entire range of energetic scales that is observed experimentally. For example, figure \ref{lowhigh} shows that the smaller-scale region of the 2-D spectrum which followed an empirically observed $\lambda_y/z \sim (\lambda_x/z)^{1/2}$ relationship is not modeled (region within the black dashed box in figure \ref{lowhigh}b). Based on the recent high-Reynolds number 2-D cross-spectrum data of \citet{deshpande2020JFM}, it was observed that $Type\,A$ eddies model the energetic scales in the logarithmic region that have a finite correlation at the wall. Their results suggest that the smaller scales unresolved by $Type\,A$ eddies are mostly wall-incoherent. This is, they do not physically extend or have a correlation with the viscous near-wall region.
Furthermore, the model also omits the contribution from the non-self-similar, very large scale attached motions \citep{baars2020a,baars2020b,yoon2020wall} that are characteristic of the superstructures in turbulent boundary layers. 

The significance of these energetic scales that are unresolved by the attached eddy model is emphasized in the recent investigations by \citet{baars2020a,baars2020b}. Their studies employ spectral-coherence based filters to decompose the measured streamwise turbulent kinetic energy into three spectral sub-components: a wall-incoherent high wavenumber component and two wall-coherent lower wavenumber components. The two wall-coherent sub-components represent the self-similar structures in the context of the attached eddy hypothesis and the non-self-similar very large scale motions. The authors report that a  $k_x^{-1}$ scaling in the 1-D streamwise spectra and the corresponding log-law in the streamwise turbulent intensity profile \citep{marusic2013logarithmic}, both indicative of self-similarity, would be masked (for $Re_\tau \lesssim 80000$) due to the overlap of the sub-component energies. Hence, at any practically encountered Reynolds number, a discussion of spectral self-similarity based on the attached eddy model is incomplete when only considering the $Type\,A$ spectral component. 

Accordingly, the objective of the current study is to extend the attached eddy model by identifying and incorporating into the conventional AEM (i) the representative energetic small-scale structures that are incoherent with the wall and (ii) the representative very large scale motions (or global modes) that are characteristic of the superstructures in turbulent boundary layers. To this end, based on the scaling of experimental 2-D spectra of $u$ for friction Reynolds numbers ranging from 2400 to 26000, the significant spectral sub-components are identified in \S\ref{model_scaling2Dspec_text}. The extension of the AEM is discussed in \S\ref{r3_extension} and the results are compared against the experiments in \S\ref{r3_comp}. Finally, a discussion on spectral self-similarity based on the extended attached eddy model is carried out in  \S\ref{r3_disc}. It is noted that throughout this paper, superscript `$+$' indicates normalization by viscous length and velocity scales, $\nu /U_\tau$ and $U_\tau$, respectively. The streamwise, spanwise and wall-normal directions are denoted by $x$, $y$ and $z$ respectively, and $u$, $v$ and $w$ denote the respective fluctuating velocity components. 

\section{Scaling of experimental 2-D spectra} 
\label{model_scaling2Dspec_text}

The extension of the AEM discussed in this paper is driven by the scaling of 2-D spectra of $u$, measured from low ($Re_\tau=2400$) to high ($Re_\tau=26000$) Reynolds numbers. These measurements were conducted in the low-$Re$ open-return boundary layer wind tunnel \citep{monty2011parametric} and the high-$Re$ boundary layer wind tunnel (HRNBLWT; \citealp{baars2016wall}) facilities, respectively, at the University of Melbourne. The 2-D correlations of $u$, and subsequently the 2-D spectra of $u$, were computed using synchronous two-point hot-wire measurements. Ref \cite{chandran2017JFM} provides full details of the experimental set-ups and measurement technique, as they are not included here for brevity.

Figures \ref{ridge}(a) and (c) show the inner- and outer-flow scalings \citep{perry1986theoretical}, respectively, of a contour of constant energy ($k_xk_y\phi_{uu}/U_\tau^2=0.15$) of the 2-D spectra as a function of $\lambda_x$ and $\lambda_y$. A constant energy contour shows the spectrum of streamwise and spanwise length scales that contribute equally to the turbulent kinetic energy. Their inner- and outer-flow scaling arguments are probed by normalizing $\lambda_x$ and $\lambda_y$ with the wall-height ($z$), and the boundary layer thickness ($\delta$), respectively. Similarly, figures \ref{ridge}(b) and (d) show the inner- and outer-flow scalings, respectively, of the energetic ridges of the 2-D spectra. The energetic ridge is computed as the maximum value of $k_xk_y\phi_{uu}/U_\tau^2$ corresponding to each streamwise wavelength, $\lambda_x$. The energetic ridge therefore indicates the aspect ratios ($\lambda_x/\lambda_y$) of the dominant energy carrying structures at a given wall-height, and can hence be used as a tool to observe geometric self-similarity \citep{del2004scaling,chandran2017JFM}.
The various scaling laws of these energetic contours and ridges, at different wall-heights and Reynolds numbers are inspected in order to prescribe the geometry and organization of the various representative eddies in the extended AEM. 

\subsection{Wall-coherent self-similar motions}
\label{Scaling_WallCohSS_text}

\begin{figure}
	\centering
	\includegraphics[trim = 0mm 0mm 0mm 0mm, clip, width=14cm]{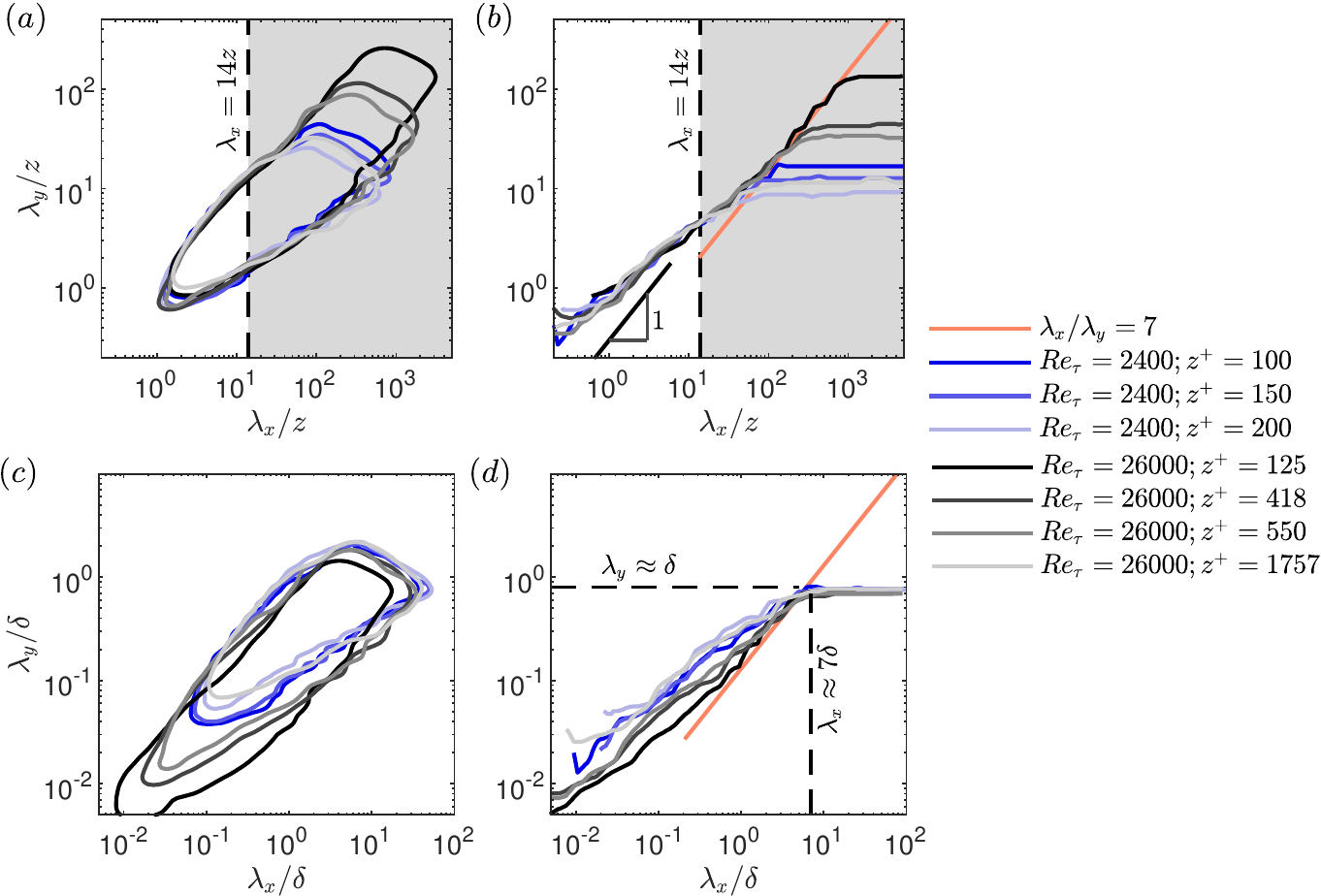} 
	\caption{(a) \& (c) Inner-flow scaling and outer-flow scaling, respectively, of the constant energy contour $k_xk_y\phi_{uu}/U_\tau^2=0.15$, and (b) \& (d) inner-flow scaling and outer-flow scaling, respectively, of the energetic ridges. The shaded region in (a) and (b) represents the wall-coherent scales as per \citet{baars2017JFM}. The boundary layer thicknesses $\delta$ at $Re_\tau=2400$ and 26000, calculated by fitting the velocity profile to the composite profile of \citet{chauhan2009criteria}, are 0.069 m and 0.337 m respectively, and the friction velocities $U_\tau$ at $Re_\tau=2400$ and 26000 are 0.545 m/s and 1.231 m/s respectively.}
	\label{ridge}
\end{figure}

Following the definition provided by \citet{baars2017JFM}, wall-coherent structures in the outer-region are portions of velocity fluctuations which correlate with the velocity fluctuations very close to the wall (or the wall-shear stress signature). \citet{baars2017JFM} isolates these wall-coherent scales from the broadband turbulence by employing an empirical filter that is based on 1-D spectral coherence (as a function of $\lambda_x$). 
They observed that the structures coherent with the wall have streamwise wavelengths $\lambda_x>14z$. These scales are represented by the dark-shaded region in figures \ref{ridge}(a) and (b). Note, that identifying the exact boundaries of the wall-coherent region in a 2-D spectrum would require a 2-D spectral-coherence based filter obtained as a function of both $\lambda_x$ and $\lambda_y$.
Hence the dark-shaded region is only an approximate reference for the wall-coherent scales, and some of the very small and the very large spanwise length scales within this region are likely incoherent with the wall. 
As discussed by \citet{chandran2017JFM}, the wavelengths of the large-scales ($\lambda_x,\lambda_y > \mathcal{O}(10z)$, hereafter referred to as the \textit{large eddy region}) tend to obey a relationship of $\lambda_y/z \sim (\lambda_x/z)^m$, where the value of $m$ approaches unity at high Reynolds numbers, or as the measurement location is moved closer to the wall (for $z>>\nu/U_\tau$). It can be observed from the inner-flow scaling of the ridges (figure \ref{ridge}(b)) that the aspect ratio of such dominant large-scale structures that tend towards self-similarity with $m=1$ is $\lambda_x/\lambda_y \approx 7$. These large-scale self-similar structures that are coherent with the wall are consistent with eddies described in Townsend's attached eddy hypothesis. Additionally, when the wavelengths are scaled in $\delta$ as shown in figure \ref{ridge}(d), the ridges collapse at $\lambda_x \approx 7\delta$ and $\lambda_y \approx \delta$ resulting in the same aspect ratio of $\lambda_x/\lambda_y \approx 7$. Therefore, the largest wall-attached self-similar structures have a characteristic `length' and `width' of roughly $7\delta$ and $1\delta$, in spectral space. These dimensions also agree with the observations of \citet{baidya2019simultaneous}, where they found that a self-similar eddy has an aspect ratio of 7 : 1 : 1 in the $x$, $y$ and $z$ directions, respectively. 

\subsection{Wall-coherent non-self-similar very large scale motions}
\label{Scaling_WallCoh_nonSS_text}

From figure \ref{ridge}(d) we observe that the large-scales nominally grow self-similarly, i.e. $\lambda_y \sim \lambda_x$, until $\lambda_x \approx 7\delta$ and $\lambda_y \approx \delta$. Beyond these limits, the ridge trends towards larger streamwise wavelengths while maintaining a constant spanwise width of $\lambda_y\approx \delta$ and with the energy dropping. At these very large scales, a good collapse of the ridges is observed with outer-flow scaling, irrespective of Reynolds number. This agrees with the findings of \citet{tomkins2005energetic}, that the structures with the largest streamwise wavelengths organize with a spanwise spacing of $\lambda_y=0.75\delta-0.9\delta$. This spanwise spacing is also consistent with the width of the anti-correlations of streamwise velocity in the spanwise direction observed by \citet{hutchins2007evidence}. 
They reported that such events tend to have long streamwise correlations; a characteristic typical of `superstructures' in boundary layer flows. It can be noted from figures \ref{ridge}(a) and (b) that these outer-scaled length scales do not follow a self-similar scaling with $z$. These structures are hence observed to be consistent with the `global' modes of \citet{del2003spectra}, and the `wall-attached non-self-similar' motions of \citet{hwang2018wall} and \citet{yoon2020wall}, that extend deep in the wall-normal direction. 
It is also noted, that even though the energetic ridges exhibit a good collapse in outer-scaling for $\lambda_x>7\delta$ (figure \ref{ridge}d), the constant energy contours (figure \ref{ridge}c) collapse only within the logarithmic region ($2.6Re_\tau^{1/2} \leq z^+ \leq 0.15Re_\tau$). The contour corresponding to $Re_\tau=26000$ and $z^+=125 (<2.6Re_\tau^{1/2})$ does not seem to follow an energetic-similarity. This trend is expected based on the findings of \citet{baars2020a}, who showed that the energy contributed by the very large-scale structures that are coherent with the wall is roughly constant for $2.6Re_\tau^{1/2} \leq z^+ \leq 0.15Re_\tau$, and reduces for $z^+<2.6Re_\tau^{1/2}$. 

\subsection{Wall-incoherent wall-scaled motions}
\label{Scaling_WallInCoh_text}

According to \citet{baars2017JFM}, wall-incoherent motions are characterized by a streamwise/wall-normal aspect ratio of $\lambda_x/z<14$ and correspond to the unshaded region in figures \ref{ridge}(a) and (b). In agreement with recent studies \citep{baars2020a,baars2020b,yoon2020wall,deSilva2020periodicity}, the contribution of wall-incoherent structures to the turbulent kinetic energy, which is the area within the 2-D spectra in the unshaded region in figure \ref{ridge}(a), appears to be significant. Even though important across the Reynolds number range studied here, the relative energy contribution of these wall-incoherent structures is observed to be more significant at low Reynolds number. 
Interestingly, as seen in figure \ref{ridge}(a), the wall-incoherent region of the 2-D spectra appears to follow a clear inner-flow scaling except for the very small Kolmogorov-type scales. This suggests the existence of wall-detached energetic motions whose characteristic lengths scale with distance from the wall. 
Moreover, the collapse of the constant energy contours suggests an invariant inner-flow scaled contribution of these wall-incoherent wall-scaled motions to the turbulent kinetic energy for all wall locations and Reynolds numbers considered here. Now, if we focus on the inner-flow scaling of the energetic ridges in this regime (figure \ref{ridge}(b)), a good collapse is observed and the ridges follow a $\lambda_y \sim \lambda_x$ behavior at these smaller scales ($\lambda_x,\lambda_y \sim \mathcal{O}(z)$), resulting in an aspect ratio of $\lambda_x/\lambda_y \approx 1$. 
Such a linear relationship at the scales $\mathcal{O}(z)$ was also reported by \citet{del2004scaling}. 
Hence, at high Reynolds number, the empirically observed $\lambda_y/z \sim (\lambda_x/z)^{1/2}$ relationship (that was predominant at low Reynolds numbers) bridges the two $\lambda_y \sim \lambda_x$ relationships observed at smaller ($\lambda_x,\lambda_y\sim\mathcal{O}(z)$) and larger ($\lambda_x,\lambda_y > \mathcal{O}(10z)$) length scales. 

\section{Extended attached eddy model} \label{r3_extension}
\begin{figure}
	\centering
	\subfigure{
		\includegraphics[trim = 0mm 0mm 0mm 0mm, clip, width=16cm]{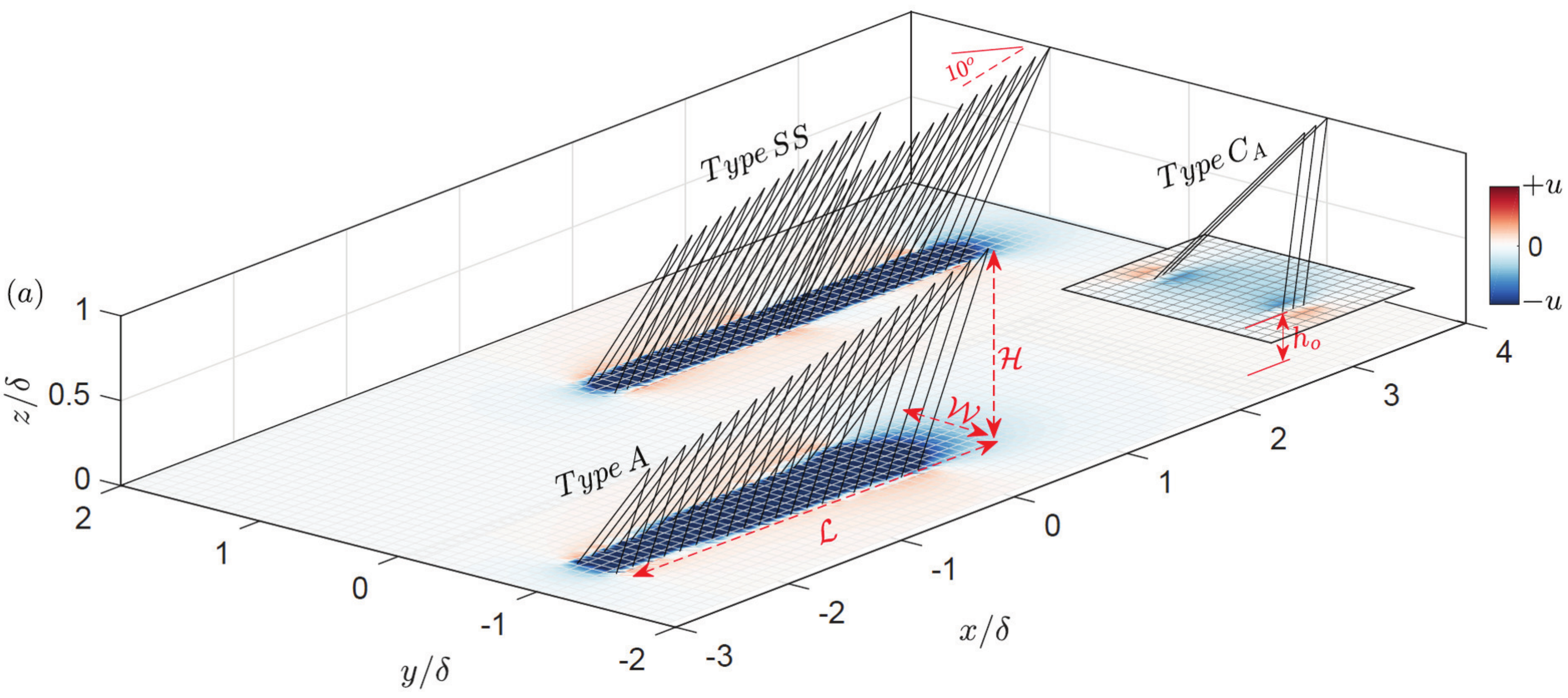}
		\label{eddy_geom}}\\
	\subfigure{
		\includegraphics[trim = 0mm 0mm 0mm 0mm, clip, width=4.8cm]{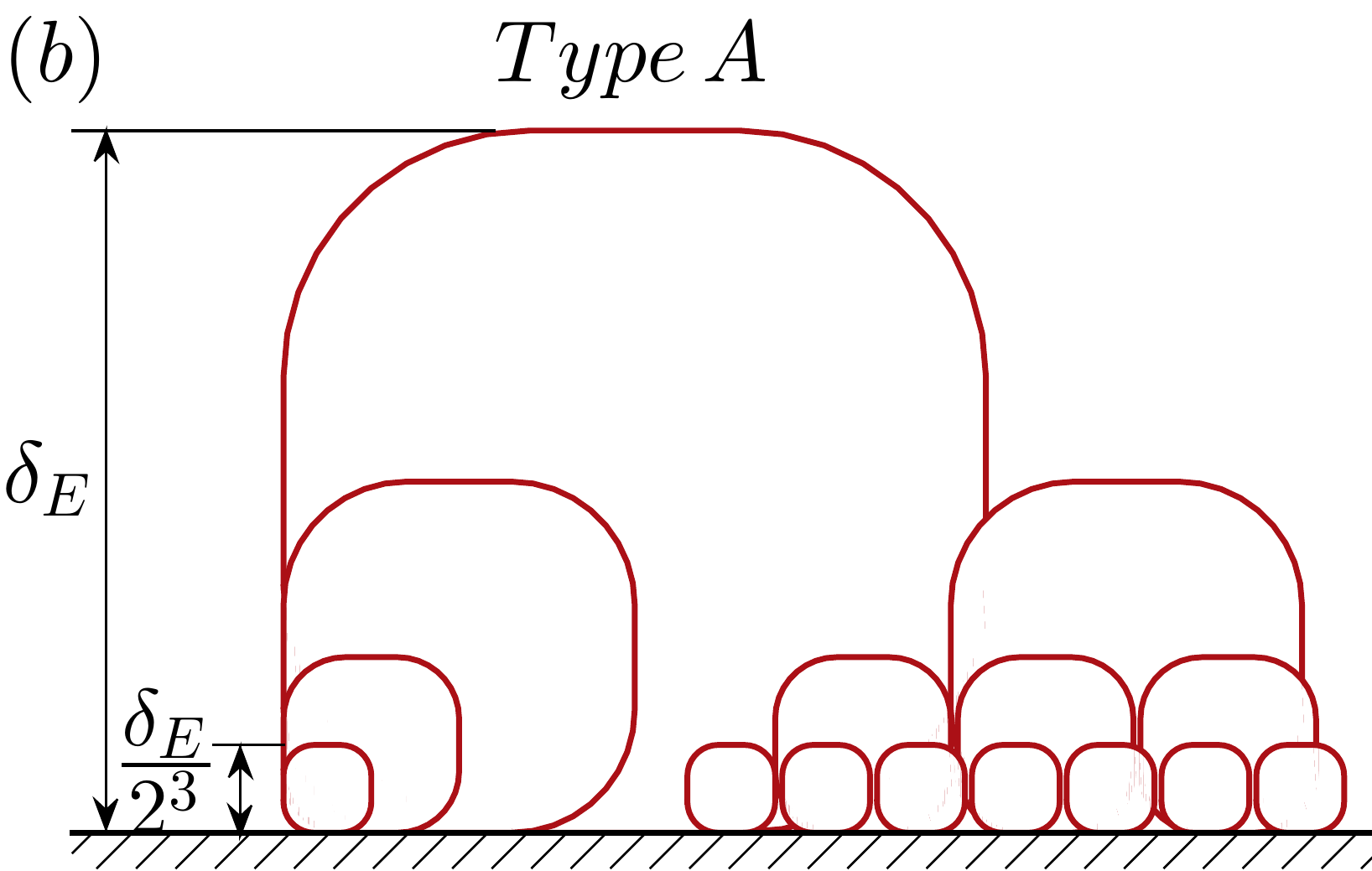}
		\label{org_A}}
	\subfigure{
		\includegraphics[trim = 0mm 0mm 0mm 0mm, clip, width=4.8cm]{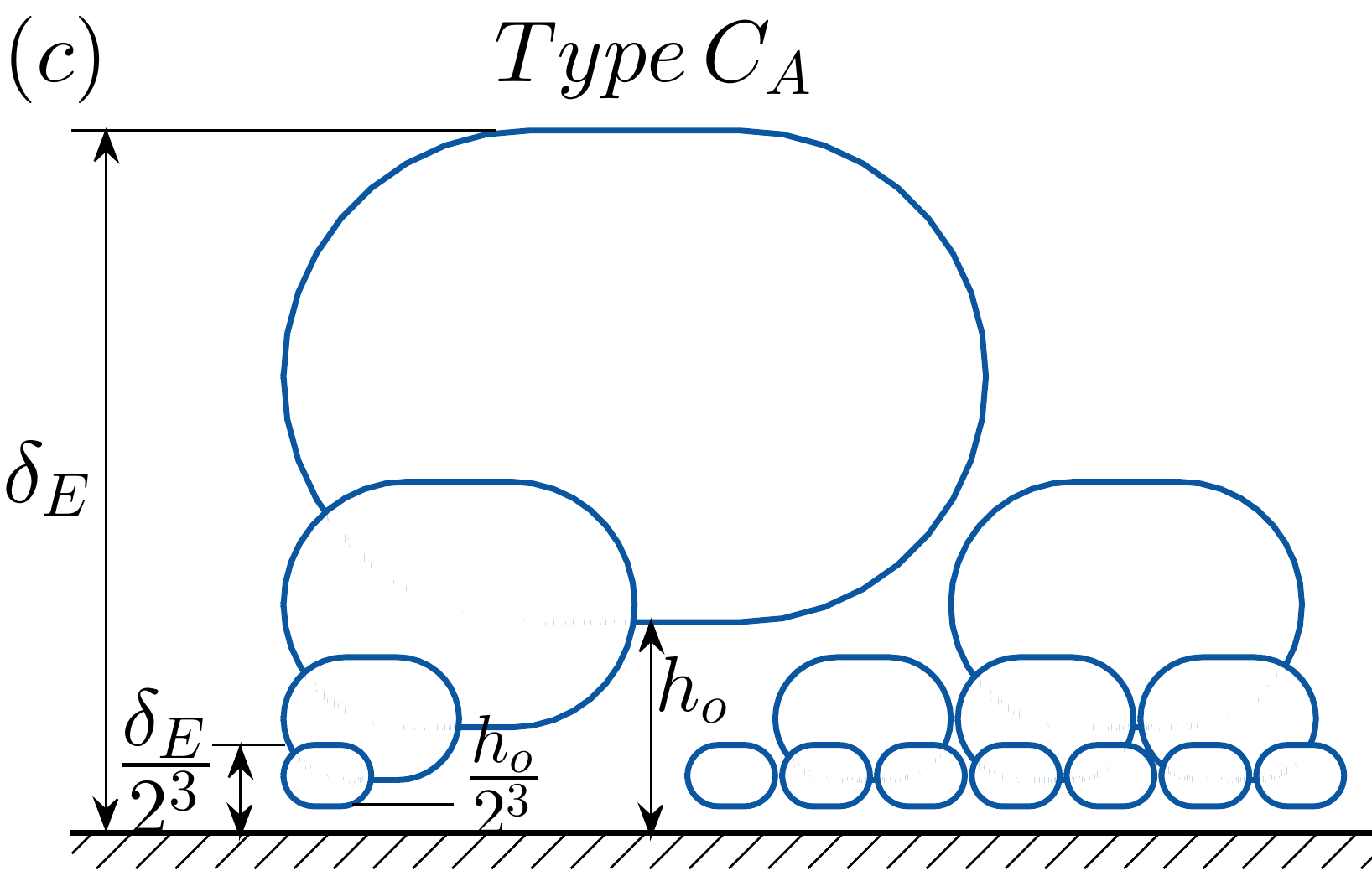}
		\label{org_CA}}
	\subfigure{
		\includegraphics[trim = 0mm 0mm 0mm 0mm, clip, width=3.79cm]{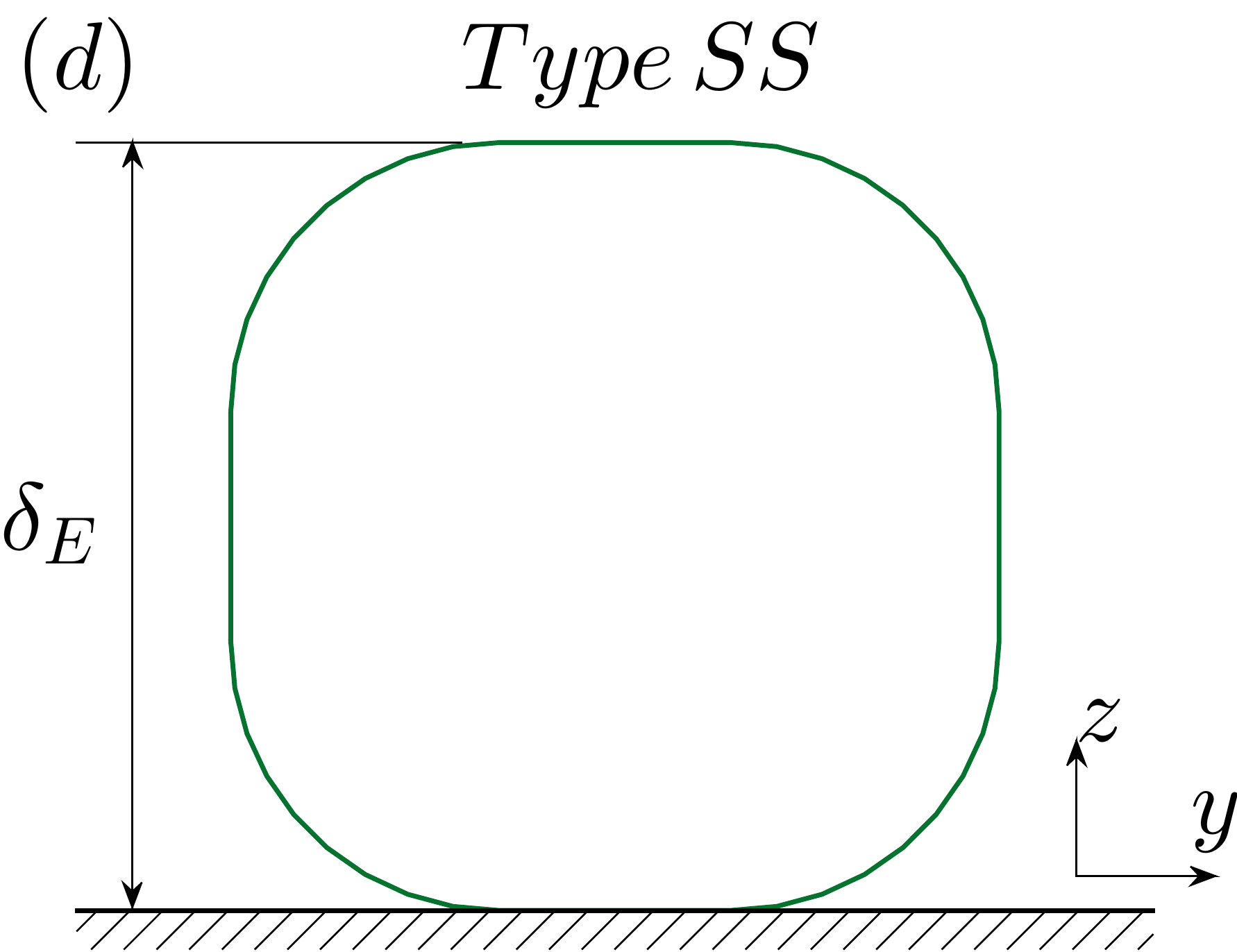}
		\label{org_SS}}
	\caption{(a) Geometry of $Type\,A$, $Type\,C_A$ and $Type\,SS$ representative packet eddies considered in the extended model with their respective contribution to the streamwise velocity and (b,c,d) schematics showing the organization of $Type\,A$, $Type\,C_A$ and $Type\,SS$ eddies respectively. The largest $Type\,C_A$ eddy ($\mathcal{H} \sim \delta_E$) is detached from the wall by $h_o$.}
	\label{eddy_geom_org}
\end{figure}

In the previous section, we identified three major contributors to the turbulent kinetic energy based on spectral-scaling 
arguments: (i) wall-coherent self-similar motions, (ii) wall-coherent very large scale motions and (iii) wall-incoherent but wall-scaled small-scale motions. Here, we attempt to  extend the attached eddy model by assigning a representative structure and an organization to each of the above identified sub-components and name them as $Type\,A$, $Type\,SS$ and $Type\,C_A$ eddies, respectively. Following prior experimental observations \citep{zhou1996autogeneration,zhou1999mechanisms,adrian2000vortex} and for simplicity, we model the $Type\,A$, $Type\,SS$ and $Type\,C_A$ representative structures using packet eddies, where hairpins at various stages of their self-similar growth are aligned in the streamwise direction \cite{dennis2011experimental,marusic2001role}. In the present study, the packets are formed by aligning `$\Lambda$-hairpins' in the streamwise direction at a growth angle of $10^{\circ} $ \cite{zhou1999mechanisms} for all three representative structures. As shown in figure \ref{eddy_geom}, the growth angle is the angle of the line connecting the heads of the first and the last hairpins in a packet. While the spacing between the hairpins are fixed for all three packet eddies, the length ($\mathcal{L}$) and width ($\mathcal{W}$) of the packets are chosen based on the scaling of the experimental 2-D spectra (\citet{monty2017measurements} has demonstrated the effect of varying the aspect ratio, $\mathcal{L}$/$\mathcal{W}$, of the representative eddy on the computed 2-D spectrum). The height ($\mathcal{H}$) of the largest eddy is maintained to be of the order of $\delta$. No other shapes for hairpins are considered in the present study as the objective is to understand the scaling of representative packet eddies rather than focusing on the exact form of individual hairpins. The organization of $Type\,A$, $Type\,C_A$ and $Type\,SS$ eddies in the boundary layer is illustrated in figures \ref{org_A} to \ref{org_SS}, respectively. The results obtained with $Type\,A$, $Type\,C_A$ and $Type\,SS$ structures are respectively color-coded using shades of red, blue and green, and the results from the composite model are represented with shades of gray.

\subsection{$Type\,A$}
$Type\,A$ eddies represent the wall-attached self-similar motions as conceptualized by Townsend \citep{townsend1980structure} and discussed in \S\ref{Scaling_WallCohSS_text} . The geometry of the representative packet ($\mathcal{L}$ and $\mathcal{W}$, as illustrated in figure \ref{eddy_geom}) is chosen such that the aspect ratio, $\mathcal{L}$/$\mathcal{W}$, is equal to the average aspect ratio of the wall-coherent self-similar motions observed in the experiments, which is $\lambda_x/\lambda_y\approx7$. The boundary layer is then populated with hierarchies of representative packet eddies that belong to different stages of their self-similar growth \citep{perry1986theoretical}. For illustrative purposes, figure \ref{org_A} represents a discrete model with four different hierarchies of $Type\,A$ eddies; the wall-normal extent of the largest and the smallest eddies in the schematic being $\delta_E (\sim \mathcal{O}(\delta))$ and $\delta_E/2^3$ respectively. The curved boxes (in figures \ref{eddy_geom_org}b-d) are illustrative of a cross-stream slice of the velocity field ($y-z$ plane) from the representative eddies and do not in any form represent the actual velocity field. Figure \ref{org_A} is similar to the physical model of \citet{perry1986theoretical} where the streamwise and spanwise extents of the representative eddies scale with distance from the wall ($z$) and their probability density is inversely proportional to $z$. However, it should be noted that unlike in figure \ref{org_A}, the actual simulation assumes a continuous hierarchy of eddies with the heights of the largest and the smallest eddies being $\mathcal{H}_L=\delta_E$ and $\mathcal{H}_S^+=100$, respectively. 
Hence, following \citet{perry1986theoretical}, the 2-D spectra resulting from this random distribution of self-similar eddies is computed as:

\begin{equation}
\frac{\phi_{uu}\left(k_x\mathcal{H},k_y\mathcal{H},\dfrac{z}{\mathcal{H}_L},\dfrac{z}{\mathcal{H}_S}\right)}{U_\tau^2} = {\displaystyle \int_{\mathcal{H}_S}^{\mathcal{H}_L}}\frac{\Phi_{uu}\left(k_x\mathcal{H},k_y\mathcal{H},\dfrac{z}{\mathcal{H}}\right)}{U_\tau^2}\mathcal{P}(\mathcal{H})\mathrm{d}(\mathcal{H}).
\label{Phi_perry_eqn}
\end{equation}
Here, $\Phi_{uu}$ is the hierarchy spectral function, which is the power spectral density of $u$ for a hierarchy of size $\mathcal{H}$ and averaged in the wall-parallel plane at a fixed $z$. $\mathcal{P}(\mathcal{H}) = 1/\mathcal{H}$ is the probability density function. Further details on computing the flow statistics from AEM can be found in \citet{perry1986theoretical} and \citet{woodcock2015statistical}. 

Figure \ref{Types_Scaling}(a) shows the 2-D spectrum of $Type\,A$ eddies at $z^+=2.6Re_\tau^{1/2}$ for $Re_\tau=26000$ (similar to figure \ref{lowhigh}(b)). The solid and dashed black lines represent $\lambda_y/z \sim \lambda_x/z$ scaling and $\lambda_y/z \sim (\lambda_x/z)^{1/2}$ relationship, respectively, as observed in experiments. As discussed in the introduction, $Type\,A$ eddies model the large scales reasonably well. Figures \ref{Types_Scaling}(b) and (c) show the inner-flow and outer-flow scaling, respectively, of contours of constant energy ($=\mathrm{max}(k_xk_y\phi_{uu}/U_\tau^2)/3$) within the logarithmic region, i.e. $2.6 Re_\tau^{-1/2}\leq z/\delta \leq 0.15$. Representative of Townsend's attached eddies, $Type\,A$ eddies follow both inner-flow and outer-flow scalings. Since the geometry of $Type\,A$ eddies is selected based on experimental data, the $Type\,A$ spectra have energy at $\lambda_x/z>14$ and the spectra grows along $\lambda_x/\lambda_y\approx7$ while moving closer to the wall. Now, considering the outer-flow scaling (when scaled in $\delta_E$) in figure \ref{Types_Scaling}(c), the large scales collapse at $\lambda_x\approx7\delta_E$ and $\lambda_y\approx\delta_E$, as in experiments (refer to \S \ref{Scaling_WallCohSS_text}). 

Figure \ref{Types_Scaling}(d) shows the wall normal profile of turbulence intensity ($\overline{u^2}^+$), which is obtained by integrating the 2-D spectrum along both the streamwise and spanwise length scales, i.e,
\begin{equation}
\overline{u^2}^+ = \int_{0}^{\infty} \int_{0}^{\infty} \frac{k_xk_y\phi_{uu}}{U_\tau^2} \, \mathrm{d}(\mathrm{ln} \, \lambda_x) \mathrm{d}(\mathrm{ln} \, \lambda_y).
\label{u2_eqn}
\end{equation}
Following from the attached eddy hypothesis, the turbulence intensity of $Type\,A$ motions decay logarithmically with increasing wall-height.

\subsection{$Type\,SS$}
\begin{figure}
	\centering
	\includegraphics[trim = 0mm 0mm 0mm 0mm, clip, width=12cm]{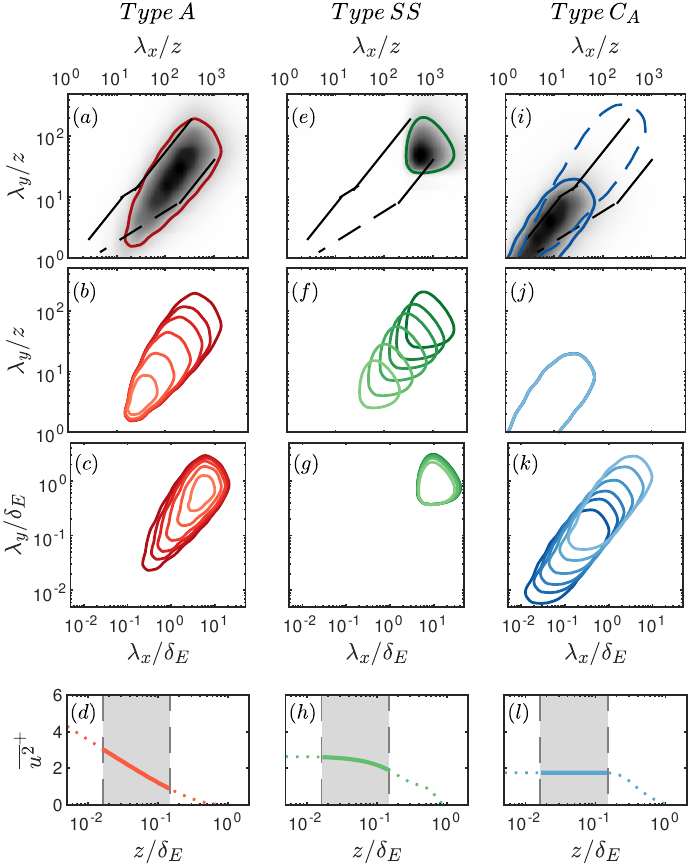}
	\caption{(a,e,i ) 2-D spectra of $u$, (b,f,j) inner-flow scaling, (c,g,k) outer-flow scaling and (d,h,l) profile of turbulence intensity of $Type\,A$, $Type\,SS$ and $Type\,C_A$ representative eddies, respectively. Line contours represent a constant energy of $\mathrm{max}(k_xk_y\phi_{uu}/U_\tau^2)/3$. Dark shade to light shade is $z/\delta=2.6 Re_\tau^{-1/2}$ to $0.15$. The blue dashed and solid line contours in (i) are from $h_o/\mathcal{H}=0$ (attached) case and $h_o/\mathcal{H}=0.15$ case respectively. The black solid and dashed lines in (a,e,i) denote the $\lambda_y/z \sim \lambda_x/z$ scaling and $\lambda_y/z \sim (\lambda_x/z)^{1/2}$ relationship, respectively.}
	\label{Types_Scaling}
\end{figure}

$Type\,SS$ eddy is representative of the wall-coherent superstructures \citep{hutchins2007evidence}, also referred to as very large scale motions \citep{kim1999very} or the `global' mode \citep{del2003spectra} that extends deep in the wall-normal direction. Following the notion of packets aligning in the streamwise direction to form longer structures \citep{kim1999very}, the $Type\,SS$ representative eddy is constructed by aligning two packets as shown in figure \ref{eddy_geom}. The length ($\mathcal{L}$) of the eddy is the total length of the two smaller packets put together. The growth angle of each packet and the spacing of hairpins within the packet are consistent with the other representative eddies. $\mathcal{L}$ and $\mathcal{W}$ are chosen such that the energy contributed by $Type\,SS$ motions is restricted only to the very large length scales (figure \ref{Types_Scaling}(e)).
Unlike the hierarchical structure of $Type\,A$ and $Type\,C_A$ eddies, as shown in figure \ref{eddy_geom_org}, the $Type\,SS$ eddy is organized as a single hierarchy with the height of the eddy $\mathcal{H}\sim\delta_E$, thereby making its contribution `global' and non-self-similar with wall-height (figure \ref{Types_Scaling}(f)).
Following the discussion in \S \ref{Scaling_WallCohSS_text}, figure \ref{Types_Scaling}(g) shows that the energy contributed by the $Type\,SS$ eddy is concentrated at the fixed outer-flow scaled wavelengths of $\lambda_x/\delta_E\approx10$ and $\lambda_y/\delta_E\approx1$, throughout the logarithmic region. 
A good collapse of the constant energy contours with outer-scaling implies a roughly constant energy contribution of $Type\,SS$ structures within the logarithmic region and consequently, 
the profile of $\overline{u^2}^+$, as plotted in figure \ref{Types_Scaling}(h), has an almost zero slope (the profile is not perfectly flat owing to the shape of the representative eddy).

\subsection{$Type\,C_A$}
\citet{marusic2019attached} describe the wall-incoherent motions as $Type\,C$ eddies that comprise of Kolmogorov-type fine scale turbulence and other wall-detached motions, some of which scale self-similarly with $z$. In the present study, we only model the inviscid subset of $Type\,C$ motions, namely $Type\,C_A$, which represent structures that are physically detached from the wall but obey a distance from the wall scaling. 
For simplicity, a packet eddy with similar growth angle and hairpin-spacing as $Type\,A$ and $Type\,SS$ is considered for $Type\,C_A$ as well (figure \ref{eddy_geom}). The organization of $Type\,C_A$ eddies in the boundary layer is illustrated in figure \ref{org_CA} with a discrete model, showing four different hierarchies. It should be noted that the organization of  $Type\,C_A$ eddies is very similar to that of $Type\,A$; the major difference being that the $Type\,C_A$ eddies are detached from the wall.  
The separation from the wall of a hierarchy of eddies of size $\mathcal{H}\sim\delta_E$ is $h_o$, and as illustrated in figure \ref{org_CA}, the wall-normal separation of any $Type\,C_A$ eddy must be a constant fraction of its wall-normal extent. This implies that the separation from the wall of $Type\,C_A$ eddies scales with $z$.  
Therefore, even when the legs of the hairpins do not extend all the way to the wall, $Type\,C_A$ eddies could be regarded as `attached' in the sense of Townsend's attached eddy hypothesis since their length scales relate to the distance from the wall \citep{marusic2019attached}. A similar organization was adopted for the $Type\,B$ eddies of \citet{perry1995wall} and \citet{marusic1995wall}, in order to model the `wake-structure' in the outer layer. 

The physical dimensions of the representative eddy, which includes $\mathcal{L}$, $\mathcal{W}$ and $h_o$, are chosen such that the energy contribution is concentrated at the smaller scales that followed the $\lambda_y\sim\lambda_x$ relationship observed in experiments (\S \ref{Scaling_WallInCoh_text}). To illustrate the effect of offsetting eddies from the wall on the 2-D energy spectrum, we first consider a case with zero separation from the wall, $h_o=0$, which is typical of $Type\,A$ organization. This is represented in figure \ref{Types_Scaling}(i) with the blue dashed line contour, which corresponds to a constant energy of $\mathrm{max}(k_xk_y\phi_{uu}/U_\tau^2)/3$. For $h_o=0$, at each wall-height $z$, the eddies with wall-normal extent $z\leq\mathcal{H}\leq\delta_E$ contributes to the turbulent kinetic energy. Hence the contour is observed to span a broad range of streamwise and spanwise length scales. Now, if we consider a finite separation from the wall for the eddies, the 2-D spectrum shrinks to smaller values of $\lambda_x/z$ and $\lambda_y/z$. The filled contour in figure \ref{Types_Scaling}(i) corresponds to a separation of $h_o/\mathcal{H}=0.15$. 
Due to the separation, at each wall-height $z$, the energy spectrum has contributions from eddies with wall-normal extent $\mathcal{H}\geq z$ and separations $h_o\leq z$. Such an organization would imply that at all wall-heights below the separation of the largest eddy, i.e, at all $z < 0.15\,\delta_E$, the energy contribution is always from a fixed number of hierarchies. Consequently, when the length scales are normalized by the distance from the wall, $z$, the constant energy contours collapse across all wall-heights within the logarithmic region (see figure \ref{Types_Scaling}(j)). This results in an invariant distribution of $\overline{u^2}^+$ for $z<0.15\,\delta_E$, as shown in figure \ref{Types_Scaling}(l). Further, it is observed from figure \ref{Types_Scaling}(k) that the energy of the $Type\,C_A$ eddies do not scale in outer units and shifts to larger length scales while moving away from the wall.

\subsection{Extended AEM}
The extended AEM constructed here is a composite model of $Type\,A$, $Type\,SS$ and $Type\,C_{A}$ eddies. The 2-D spectra from the AEM is hence computed as,

\begin{equation}
k_xk_y\phi_{uu,COMP}^+ = \frac{(k_xk_y\phi_{uu,A}+W_{SS}\times k_xk_y\phi_{uu,SS}+W_{CA}\times k_xk_y\phi_{uu,CA})}{U_{\tau,COMP}^2} 
\label{kkcomp_eqn}
\end{equation}
where,
\begin{equation}
U_{\tau,COMP}^2 = \mathrm{max}(-\overline{uw}_{COMP})
\label{Utau_eqn}
\end{equation} 
and
\begin{equation}
\begin{split}
\overline{uw}_{COMP} = \int\int_{-\infty}^{\infty}\bigg\lgroup(k_xk_y\phi_{uw,A}+W_{SS}\times k_xk_y\phi_{uw,SS}+W_{CA}\times k_xk_y\phi_{uw,CA})\bigg\rgroup\\
\mathrm{d}(\mathrm{ln} \, \lambda_x) \mathrm{d}(\mathrm{ln} \, \lambda_y).
\label{uw_eqn}
\end{split}
\end{equation}    
Here, $k_xk_y\phi_{uu,A}$, $k_xk_y\phi_{uu,SS}$, and $k_xk_y\phi_{uu,C_A}$ represent the 2-D spectrum of $u$ from $Type\,A$, $Type\,SS$ and $Type\,C_A$ eddies respectively. $W_{SS}$ and $W_{CA}$ are the relative weightings for the energy contributions from $Type\,SS$ and $Type\,C_A$ eddies respectively, in relation to the energy contribution from $Type\,A$ eddies. Changing the values of $W_{SS}$ and $W_{CA}$ will change the shape of the composite 2-D spectrum and their values are chosen arbitrarily to match the composite 2-D spectrum with experiments (discussed in \S \ref{AEMresults_spectra of u_text}). We note, the objective of introducing these weightings is not to match the magnitude of $k_xk_y\phi_{uu}/U_\tau^2$ with experimental values, but to get the distribution of energy among the right length scales, i.e, to get the correct shape of the 2-D spectrum. The composite friction velocity, $U_{\tau,COMP}$ is computed by forcing the inner-normalized peak Reynolds shear stress in the logarithmic region to be unity \citep{buschmann2009near}, i.e, peak $-\overline{uw}^+_{COMP}= \mathrm{max}(-\overline{uw}_{COMP}/U_{\tau,COMP}^2)=1$. As represented in equation \ref{uw_eqn}, $\overline{uw}_{COMP}$ is computed by integrating the composite 2-D $uw-$spectrum across $\lambda_x$ and $\lambda_y$. The same weightings ($W_{SS}$ and $W_{CA}$) for the $Type\,SS$ and $Type\,C_A$ contributions, as in equation \ref{kkcomp_eqn}, are used in equation \ref{uw_eqn}.

\section{Results from the extended AEM} \label{r3_comp}

\subsection{Spectra of $u$} \label{AEMresults_spectra of u_text}
\begin{figure}
	\centering
	\includegraphics[trim = 0mm 0mm 0mm 0mm, clip, width=12cm]{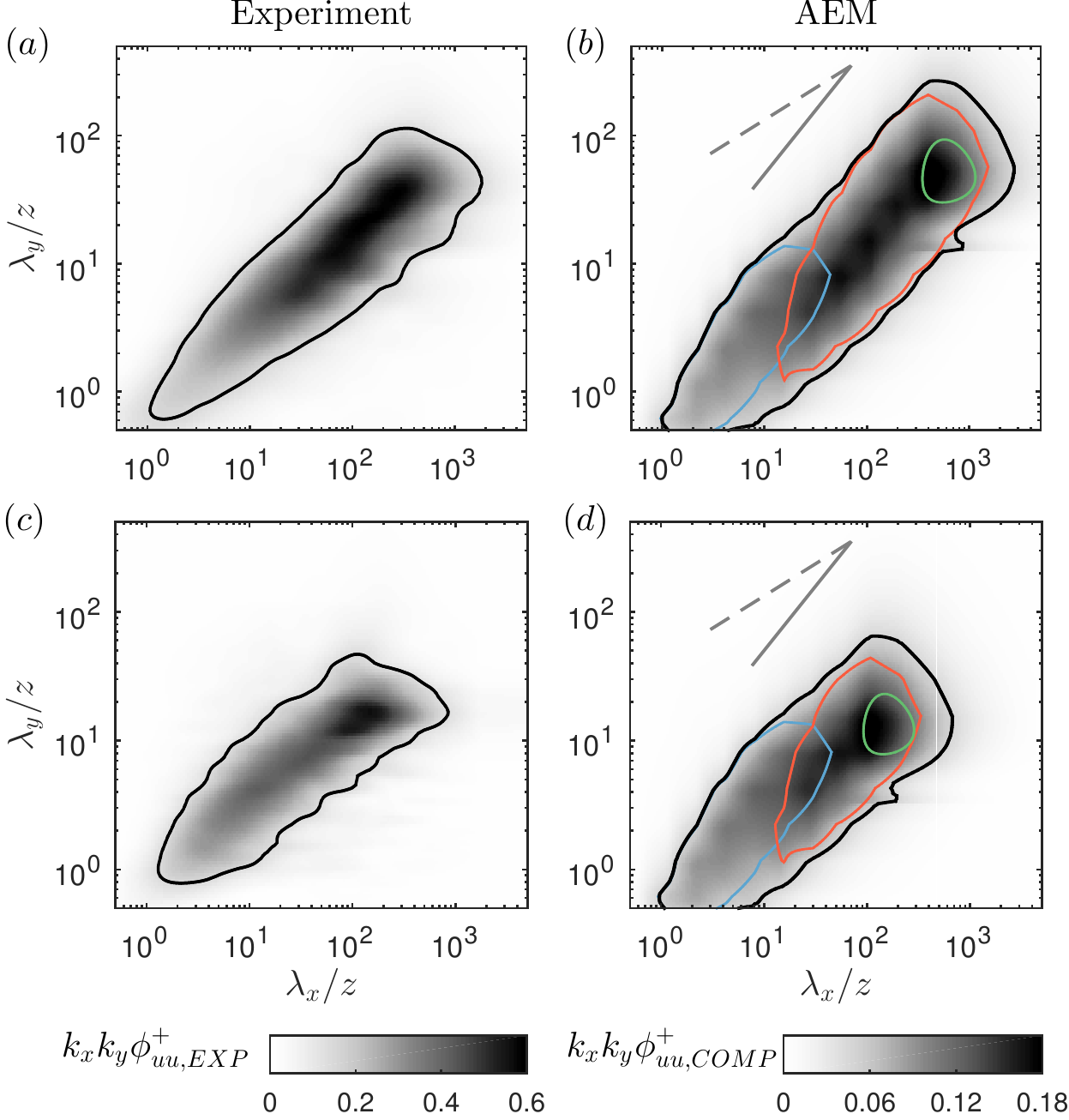}
	\caption{Comparison of 2-D spectra of $u$ from the extended AEM with experiments at $z^+=2.6Re_\tau^{1/2}$ for (a,b) $Re_\tau\approx26000$ and (c,d) $Re_\tau\approx2400$. The line contour represents $\mathrm{max}(k_xk_y\phi_{uu}^+)/4$. In (b,d) black, red, green and blue contours represent composite, $Type\,A$, $Type\,SS$ and $Type\,C_A$ spectra respectively and the gray solid and dashed lines are the references for $\lambda_y/z \sim \lambda_x/z$ scaling and $\lambda_y/z \sim (\lambda_x/z)^{1/2}$ relationship, respectively.}
	\label{comp_u}
\end{figure}

Figure \ref{comp_u}(a) and (b) show the 2-D spectra of $u$ at $Re_\tau\approx26000$ and $z^+=2.6Re_\tau^{1/2}$ from the experiments and the extended model, respectively. The line contours represent $\mathrm{max}(k_xk_y\phi_{uu}^+/4)$. Additionally, in figure \ref{comp_u}(b), the contributions of $Type\,A$, $Type\,SS$ and $Type\,C_A$ eddies to the composite spectra is shown with red, green and blue colored contours, respectively. The values of $W_{SS}$ and $W_{CA}$ are chosen to be 0.4 and 14 respectively, in order to match the shape of the composite 2-D spectrum with experiments. These values for $W_{SS}$ and $W_{CA}$ are fixed for the representative structures considered in the present study and do not vary with respect to Reynolds numbers, wall-locations or the components of velocity. However, these values are specific for the current representative eddies (figure \ref{eddy_geom}) and would change with the shape of the hairpin, spacing between hairpins in a packet, strength of the vortex rods etc. 

It is observed from figures \ref{comp_u}(a) and (b) that the composite spectra obtained with the extended AEM, captures the major trends observed in the high-$Re$ experimental 2-D spectra. While the conventional AEM that comprises $Type\,A$ eddies alone (red contour in figure \ref{comp_u}(b)), represents only the large-scales in the 2-D spectra, the extended model predicts a broader range of length scales, from $\mathcal{O}(z)$ to $\mathcal{O}(10\delta)$. The $\lambda_y/z \sim \lambda_x/z$ behavior observed at the smaller length scales in experimental spectra and \citet{del2004scaling}, is now captured using $Type\,C_A$ eddies. The shape of the 2-D spectra at very large length scales is also comparable with experiments due to $Type\,SS$ contributions. Interestingly, the length scales, where $Type\,C_A$ and $Type\,A$ spectra overlap, is observed to follow a near-square-root ($\lambda_y/z \sim (\lambda_x/z)^{1/2}$) behavior. This agrees with experiments where the square-root relationship was observed to bridge the two $\lambda_y \sim \lambda_x$ relationships observed at smaller ($\lambda_x,\lambda_y\sim\mathcal{O}(z)$) and larger ($\lambda_x,\lambda_y > \mathcal{O}(10z)$) length scales, as discussed in \S \ref{model_scaling2Dspec_text}.

Contrary to high-$Re$ spectra, the square-root relationship is predominant at low-$Re$, even at larger scales. Hence, the conventional AEM with $Type\,A$ eddies alone does not predict the large-scales at low-$Re$ (red line contour in figure \ref{comp_u}d). 
Since the extended AEM incorporates the low-$Re$ characteristics with $Type\,C_A$ contributions, a better prediction is observed even at low Reynolds numbers, as shown in figure \ref{comp_u}(d). 
It is observed, that at low Reynolds number ($Re_\tau\approx2400$), the range of length scales for $Type\,A$ is narrow compared to $Re_\tau\approx26000$ and hence the scale separation between $Type\,C_A$ and $Type\,SS$ is less. Therefore, the $Type\,A$, $Type\,C_A$ and $Type\,SS$ energy spectra overlap at the larger length scales, resulting in a trend similar to the $\lambda_y/z \sim (\lambda_x/z)^{1/2}$ relationship, as observed in the experimental low-Reynolds number spectrum (figures \ref{comp_u}(c) and \ref{comp_u}(d)). The weaker $Type\,A$ contribution at low-Reynolds numbers prohibits a transition of this square-root relationship to a $\lambda_y/z \sim \lambda_x/z$ trend observed at high Reynolds numbers. 

\subsubsection{Inner-flow scaling of 2-D spectra of $u$} \label{InnserScaling_spectra of u_text}

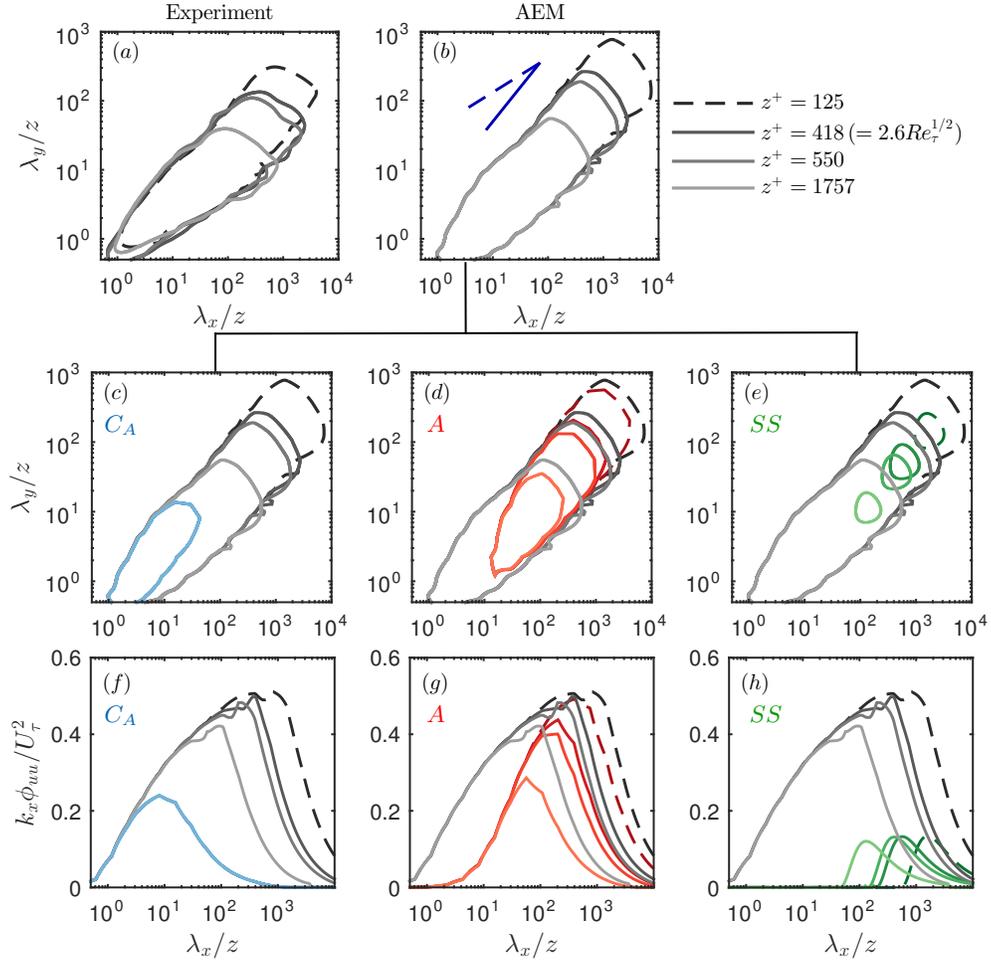
\begin{figure}
	\centering
	\includegraphics[trim = 0mm 0mm 0mm 0mm, clip, width=13cm]{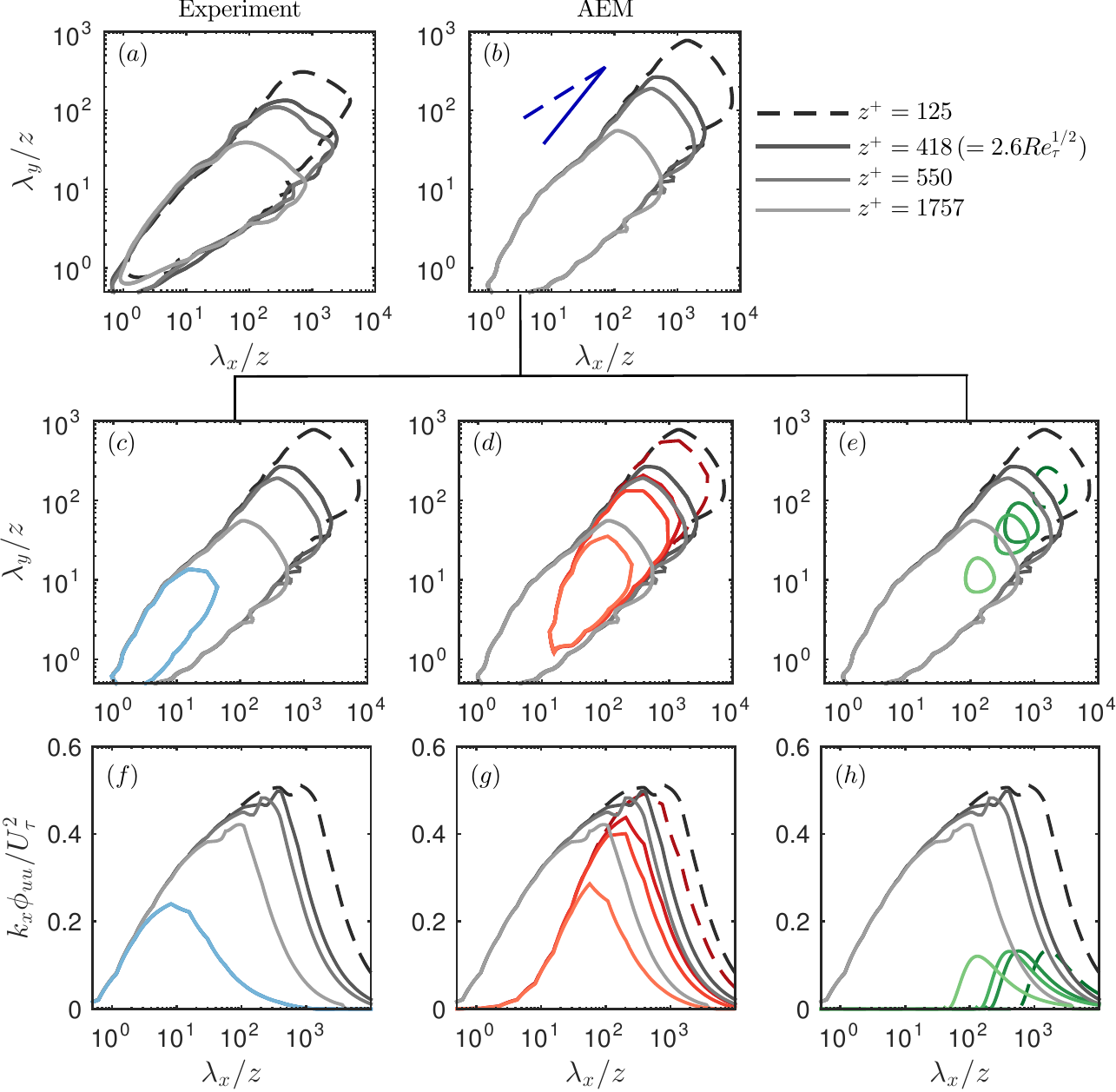}
	\put(-90,90){{\color{green2}$SS$}}
	\put(-90,200){{\color{green2}$SS$}}
	\put(-212,90){{\color{red}$A$}}
	\put(-212,200){{\color{red}$A$}}
	\put(-334,90){{\color{blue2}$C_A$}}
	\put(-334,200){{\color{blue2}$C_A$}}
	\caption{Inner-flow scaling of 2-D spectra of $u$: (a) Experiments at $Re_\tau\approx26000$, (b) Composite spectra from the extended AEM at $Re_\tau\approx26000$, (c,d,e) highlighting $Type\,C_A$ (blue), $Type\,A$ (red) and $Type\,SS$ (green) contributions to the composite 2-D spectra and (f,g,h) highlighting $Type\,C_A$, $Type\,A$ and $Type\,SS$ contributions to the composite 1-D streamwise spectra. The line contours represent a constant energy of $\mathrm{max}(k_xk_y\phi_{uu}^+|_{z^+=125})/4$. The blue solid and dashed lines in (b) are the references for $\lambda_y/z \sim \lambda_x/z$ scaling and $\lambda_y/z \sim (\lambda_x/z)^{1/2}$ relationship, respectively.}
	\label{comp_u_zscale}
\end{figure}

The inner-flow scaling ($z$-scaling) of the 2-D spectra of $u$, obtained from the extended AEM, at $Re_\tau \approx 26000$ and computed at different wall-heights is shown in figure \ref{comp_u_zscale}(b). The results are compared against the experimental spectra (figure \ref{comp_u_zscale}a) at matched Reynolds number and wall-heights. 
The comparison reveals a good agreement between figures \ref{comp_u_zscale}(a) and (b) with the 2-D spectra showing a good collapse at the smaller streamwise and spanwise wavelengths ($\mathcal{O}(z)$ to $\mathcal{O}(10z)$), for the wall-heights considered. 
Figures \ref{comp_u_zscale}(c) and (d) show that this collapse is a result of the perfect $z$-scaling of the $Type\,C_A$ spectra and the small-scale-end of $Type\,A$ spectra. The Reynolds number invariance and hence the low-$Re$ trend at the small-scales is effected by the $Type\,C_A$ contribution, which follows a $\lambda_y/z \sim \lambda_x/z$ relationship. Within the region of collapse, this linear growth is observed to transition towards a square-root $\lambda_y \sim \lambda_x^{1/2}$ behavior when there is an overlap between $Type\,C_A$ and $Type\,A$ energies. 
Now, at wavelengths larger than $(\lambda_x,\lambda_y)\sim\mathcal{O}(10z)$ in the \textit{large eddy region}, the spectra  deviates from the scaling and trends towards the $\lambda_y/z \sim (\lambda_x/z)$ relationship, as observed in experiments. This transition towards a linear relationship in the \textit{large eddy region} is dictated by $Type\,A$ energy (figure \ref{comp_u_zscale}(d)) and therefore the $\lambda_y/z \sim (\lambda_x/z)$ scaling is more pronounced, due to the increasing contribution of $Type\,A$, as we move closer to the wall (or increasing $Re_\tau$ at $z>>\nu/U_\tau$). We note that a peel-off from the $z$-scaling at the very small scales is not observed in the AEM since the high-frequency Kolmogorov-type motions are not modeled.

Since 1-D streamwise spectra has been a popular tool to observe self-similarity, the composite 1-D streamwise spectra highlighting the contributions from  $Type\,C_A$, $Type\,A$ and $Type\,SS$ eddies is shown as figures \ref{comp_u_zscale}(f,g,h) respectively. Figures \ref{comp_u_zscale}(f,g,h) are obtained by integrating figures \ref{comp_u_zscale}(c,d,e) respectively, across the whole range of spanwise length scales $\lambda_y$ as:
\begin{equation}
k_x\phi_{uu}^+(k_x)  =  \int_{0}^{\infty} k_xk_y\phi_{uu}^+(k_x,k_y)\,\mathrm{d}(\mathrm{ln} \, \lambda_y).
\label{eq_1dxspectra}
\end{equation}
Figures \ref{comp_u_zscale}(f,g,h) are qualitatively comparable with the triple-decomposed spectra of \citet{baars2020a} (figure 15 (f,d,b) respectively in their paper) where the decomposition technique used empirically obtained coherence based filters. As observed by \citet{baars2020a}, the maxima of the wall-incoherent small-scale ($Type\,C_A$) energy is located at $\lambda_x \sim \mathcal{O}(10z)$. Additionally, beyond $\lambda_x \sim \mathcal{O}(10z)$, the $Type\,A$ spectra is observed to ramp-up with its amplitude increasing with decreasing wall-height. The ramp-up of the $Type\,A$ spectra at its small-scale end appears to scale with $z$, which is in agreement with the empirical observation of \citet{baars2020a}. As indicated in figures \ref{comp_u_zscale}(e) and (h), $Type\,SS$ motions do not contribute to the wall-scaling of the composite spectra.

\subsubsection{Outer-flow scaling of 2-D spectra of $u$}
\begin{figure}
	\centering
	\includegraphics[trim = 0mm 0mm 0mm 0mm, clip, width=13cm]{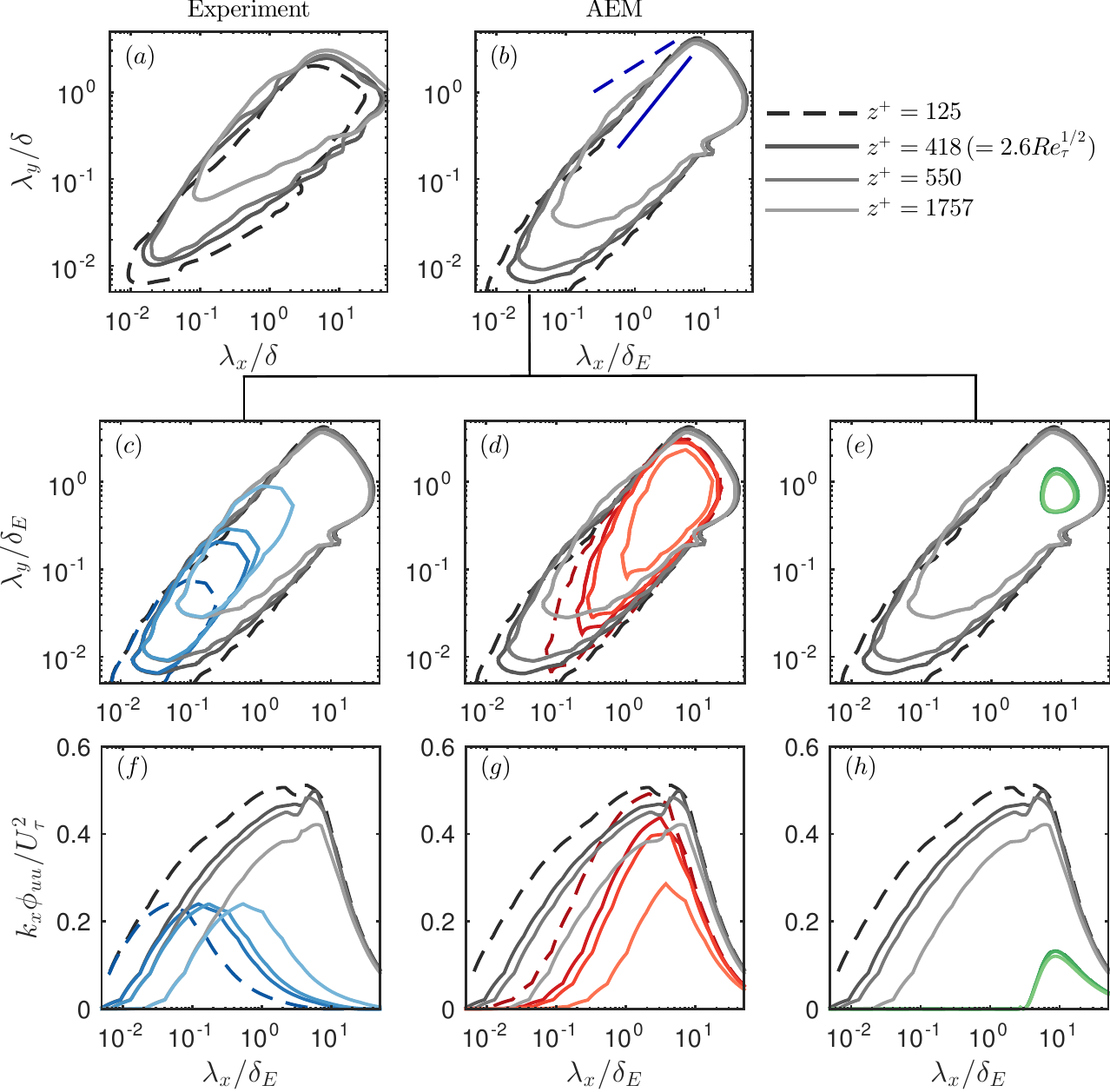}
	\put(-87,93){{\color{green2}$SS$}}
	\put(-87,200){{\color{green2}$SS$}}
	\put(-209,93){{\color{red}$A$}}
	\put(-209,200){{\color{red}$A$}}
	\put(-331,93){{\color{blue2}$C_A$}}
	\put(-331,200){{\color{blue2}$C_A$}}
	\caption{Outer-flow scaling of 2-D spectra of $u$ when the wavelengths are normalized by the boundary layer thickness $\delta$ (in experiments) or the height of the largest eddy $\delta_E$ (in AEM). Details of the plots are the same as in figure \ref{comp_u_zscale}.}
	\label{comp_u_dscale}
\end{figure}

The outer-flow scaling ($\delta$-scaling) of the composite 2-D spectra of $u$, obtained from the extended AEM, at $Re_\tau \approx 26000$ and computed at different wall-heights is shown in figure \ref{comp_u_dscale}(b) and are compared against experiments (figure \ref{comp_u_dscale}(a)). As observed in experimental data, the composite 2-D spectra show a good collapse at scales larger than $\lambda_x\approx 7\delta_E$ and $\lambda_y\approx\delta_E$ for all wall-heights considered. As observed in figures \ref{comp_u_dscale}(d) and (e), this collapse is due to the $\delta$-scaled contributions from the $Type\,SS$ eddies and the large-scale end of $Type\,A$.
Now, as observed in experiments, for $\lambda_x<7\delta_E$ in the \textit{large eddy region}, the constant energy contour deviates from a perfect $\delta$-scaling while following the relationship of $\lambda_y/\delta \sim (\lambda_x/\delta)^m$. The value of $m$ is observed to transition from 0.5 to 1 (represented by dashed and solid blue lines respectively in figure \ref{comp_u_dscale}b) as we move closer to the wall.  From figures \ref{comp_u_dscale}(c) to (e), we see that this trend towards $m=1$ with decreasing $z$ is due to the increased $Type\,A$ contribution and thus the increased scale separation between $Type\,SS$ and $Type\,C_A$ energies. 

The $\delta$-scaled composite 1-D streamwise spectra highlighting the contributions from  $Type\,C_A$, $Type\,A$ and $Type\,SS$ eddies are shown as figures \ref{comp_u_dscale}(f,g,h) respectively. Figures \ref{comp_u_dscale}(f,g,h) are comparable with the triple-decomposed spectra of \citet{baars2020a} (figure 15 (e,c,a) respectively in their paper). As observed by \citet{baars2020a}, the maxima of the wall-coherent very large scale ($Type\,SS$) energy is located at $\lambda_x \sim \mathcal{O}(10\delta_E)$. The roll-off at the large-scale end of $Type\,A$ spectra is observed to follow $\delta$-scaling in agreement with the empirical observation of \citet{baars2020a} for $2.6Re_\tau^{1/2}\leq z^+\leq 0.15\delta^+$.  
As indicated in figures \ref{comp_u_dscale}(c) and (f), $Type\,C_A$ motions do not contribute to the outer-flow scaling of the composite spectra.

\subsection{Spectra of $v$ and $w$}

\begin{figure}
	\centering
	\includegraphics[trim = 0mm 0mm 0mm 0mm, clip, width=15cm]{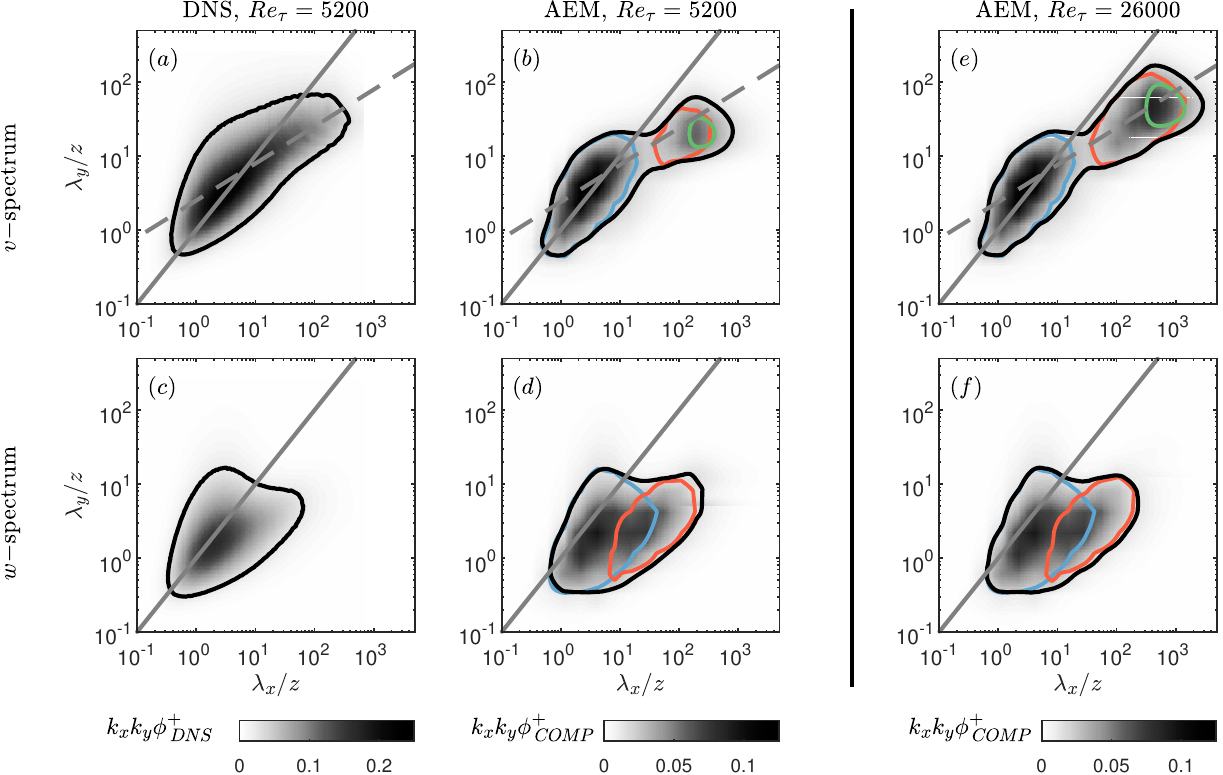}
	\caption{(a-d) Comparison of the extended AEM with DNS of \citet{lee2015moser} at $Re_\tau=5200$ and $z^+=2.6Re_\tau^{1/2}$; (a,b) spectra of spanwise velocity ($v$) and (c,d) spectra of wall-normal velocity ($w$). (e,f) Spectra of $v$ and $w$, respectively, as predicted by the extended AEM for $Re_\tau=26000$. The line contours in all panels represent a constant energy of $\mathrm{max}(k_xk_y\phi^+)/6$. In (b,d,e,f) black, red, green and blue contours represent composite, $Type\,A$, $Type\,SS$ and $Type\,C_A$ spectra respectively. The gray solid and dashed lines denote $\lambda_y/z=\lambda_x/z$ and $\lambda_y/z \sim (\lambda_x/z)^{1/2}$ respectively.}
	\label{comp_vw}
\end{figure}
In the present study, the extension to the AEM was driven by scaling of the 2-D spectra of the streamwise velocity $u$ alone, since only $u$ data are available at high Reynolds number. We can now assess whether the extended AEM also provides better predictions for the spectra of the spanwise ($v$) and the wall-normal ($w$) velocity components.  
The composite 2-D spectra of $v$ and $w$ are computed from the model similar to the computation of the spectra of $u$ (equation \ref{kkcomp_eqn}) with the values of $W_{CA}$ and $W_{SS}$ remaining the same. The results from the model are compared with the DNS of \citet{lee2015moser} at $Re_\tau=5200$ in figure \ref{comp_vw}(a-d). We chose this dataset as it is the highest $Re$ data available for the 2-D spectra of $v$ and $w$. As in figure \ref{comp_vw}(a-d), spectra of $v$ and $w$ from the extended AEM show good agreement with DNS. 
It is seen from the DNS data that the dominant streamwise and spanwise modes in both $v-$ and $w-$spectra are $\mathcal{O}(z)$. These energetic modes are represented in the model with the major contribution from $Type\,C_A$ eddies for the $Re_\tau=5200$ case. As shown in figures \ref{comp_vw}(b) and (d), a model with $Type\,A$ eddies alone (red line contours) represent the dominant modes at much larger length scales in the $v-$ and $w-$spectra (as also observed by \citet{baidya2017distance}) and misses a large contribution to the overall energy. 
It is observed that the energy not represented by $Type\,A$ eddies, or the conventional AEM, is more significant for the $v-$ and $w-$spectra in comparison to the spectra of $u$.

\begin{figure}
	\centering
	\includegraphics[trim = 0mm 0mm 0mm 0mm, clip, width=17cm]{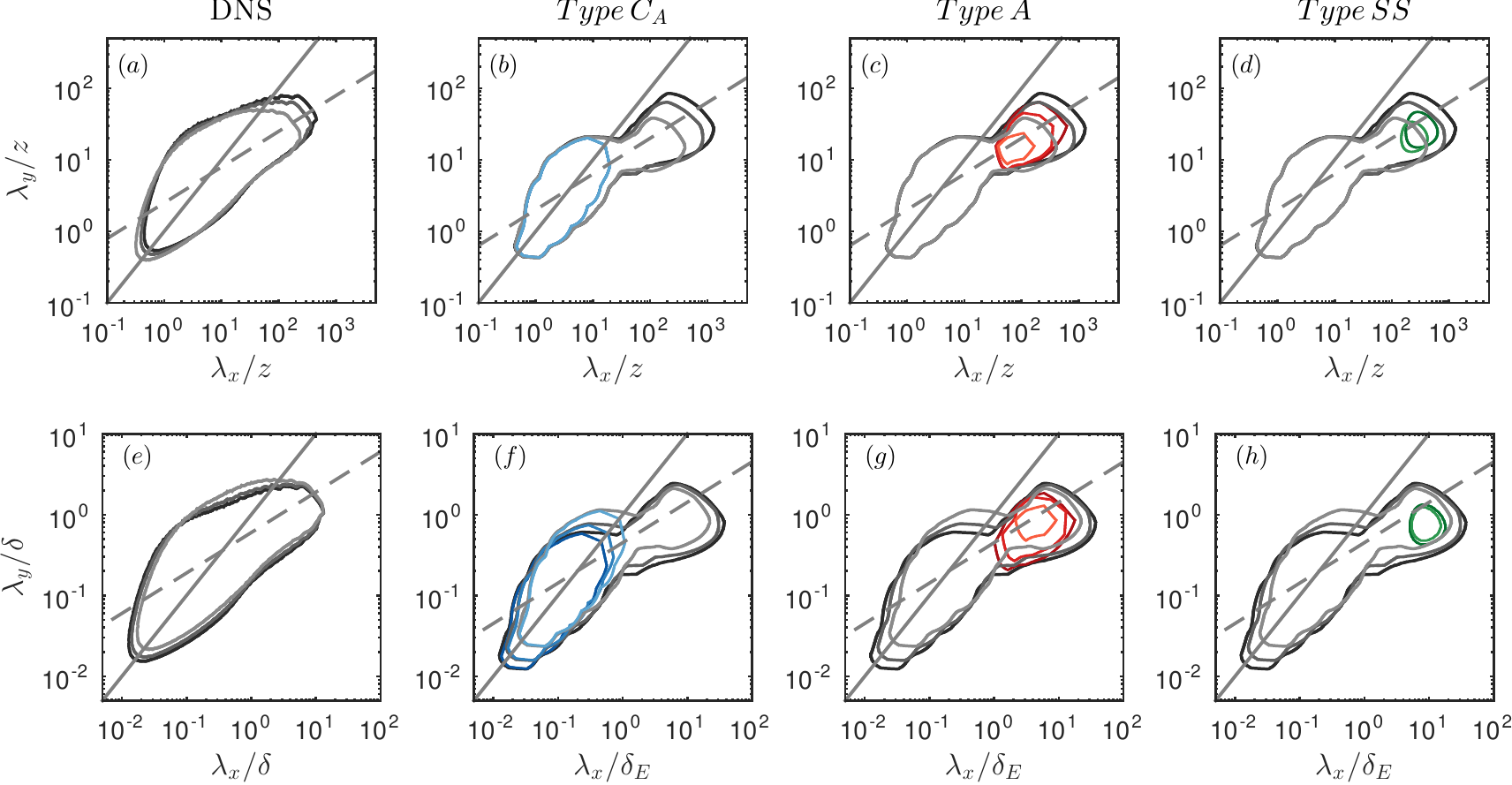}
	\caption{Inner-flow scaling of 2-D spectra of $v$ from (a) the DNS of \citet{lee2015moser} and (b,c,d) extended AEM highlighting the contributions from  $Type\,C_A$ (blue), $Type\,A$ (red) and $Type\,SS$ (green) to the $v-$spectra respectively, at $z^+=150,\,2.6Re_\tau^{1/2}$ and $3.9Re_\tau^{1/2}$ (dark to light shade respectively). (e-h) Similar plots in outer-flow scaling. The line contours represent a constant energy of $\mathrm{max}(k_xk_y\phi_{vv}^+|_{z^+=150})/6$ . The gray solid and dashed lines denote $\lambda_y=\lambda_x$ and $\lambda_y \sim (\lambda_x)^{1/2}$ respectively.}
	\label{vspectra_scaling}
\end{figure}

At low Reynolds number ($Re_\tau=2000$), \citet{jimenez2008turbulent} reported from their DNS of a  channel flow that the energetic ridge of the 2-D spectra of $v$ and $w$ follow a $\lambda_y/z=\lambda_x/z$ relationship in the log region.  However, from the data at higher Reynolds numbers ($Re_\tau=5200$, DNS), we see that in the $v-$ spectra, above scales $\mathcal{O}(10z)$, the $\lambda_y/z=\lambda_x/z$ relationship transitions to a square-root relationship of $\lambda_y/z \sim (\lambda_x/z)^{1/2}$, similar to the trend observed in the 2-D spectra of $u$. This trend is expected as $u-$ and $v-$ spectra follow similar scaling laws \citep{perry1986theoretical}. 
To understand this better, the inner-flow and outer-flow scaling of the composite 2-D spectra of $v$ is plotted in figure \ref{vspectra_scaling} highlighting the contributions of $Type\,C_A$, $Type\,A$ and $Type\,SS$ eddies. The predominant low-$Re$ trend of $\lambda_y/z=\lambda_x/z$ is due to the $Type\,C_A$ contribution that scales with $z$. As observed for the $u-$spectra, the transition to a square-root relation appears to be at scales ($\mathcal{O}(10z)$) where the $Type\,C_A$ and the $Type\,A$ energies overlap. At scales larger than $\mathcal{O}(10z)$, the shape of the 2-D spectra is dictated by $Type\,A$ energy which seem to gradually transition towards a $\lambda_y/z\sim\lambda_x/z$ relationship. 
However, the current Reynolds number ($Re_\tau=5200$) is not high enough for this linear trend to be conspicuous. Figure \ref{comp_vw}(e,f) show the predictions of $v-$ and $w-$spectra from the extended AEM at $Re_\tau \approx 26000$. Similar to the \textit{large eddy region} in the spectra of $u$, the large scales ($>\mathcal{O}(10z)$) in a constant energy region of the $v-$ spectrum appear to begin to follow the  $\lambda_y/z \sim \lambda_x/z$ scaling, indicating self-similarity.
A validation of this self-similar trend requires the measurement of the 2-D $v-$ spectra at high Reynolds numbers.

\begin{figure}
	\centering
	\includegraphics[trim = 0mm 0mm 0mm 0mm, clip, width=16cm]{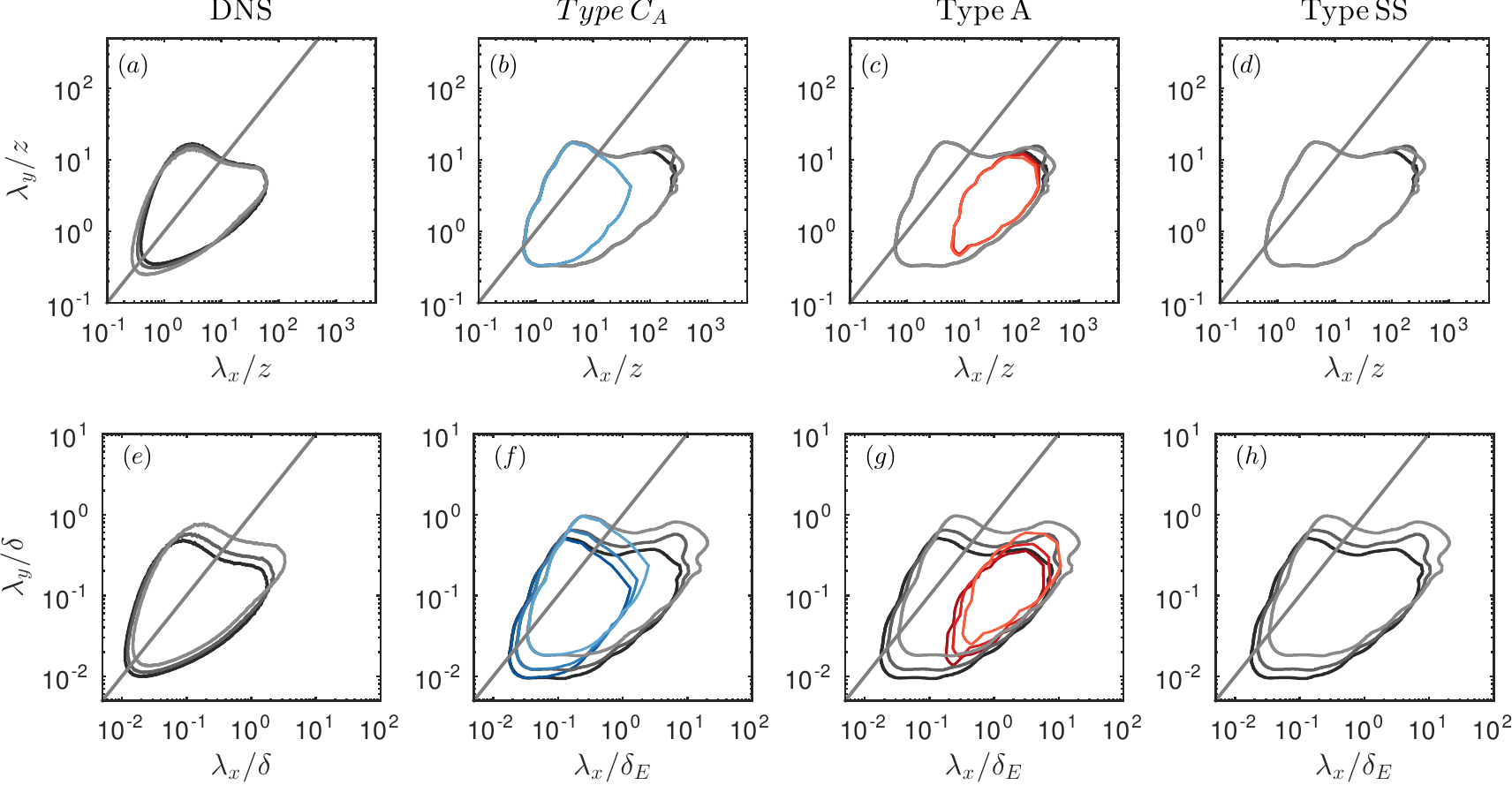}
	\caption{Inner-flow scaling of 2-D spectra of $w$ from (a) the DNS of \citet{lee2015moser} and (b,c,d) extended AEM highlighting the contributions from  $Type\,C_A$ (blue), $Type\,A$ (red) and $Type\,SS$ (green) to the $w-$spectra respectively, at $z^+=150,\,2.6Re_\tau^{1/2}$ and $3.9Re_\tau^{1/2}$ (dark to light shade respectively). (e-h) Similar plots in outer-flow scaling. The line contours represent a constant energy of $\mathrm{max}(k_xk_y\phi_{ww}^+|_{z^+=150})/6$ . The gray solid and dashed lines denote $\lambda_y=\lambda_x$ and $\lambda_y \sim (\lambda_x)^{1/2}$ respectively.}
	\label{wspectra_scaling}
\end{figure}

Unlike the $u$ and $v$ components, at a particular wall-height $z$, only those eddies with heights $\mathcal{H} \sim z$ contribute to $w-$ spectra. Hence, as shown in figure \ref{wspectra_scaling}, the $w-$ spectra follows a perfect inner-flow scaling \citep{perry1986theoretical,baidya2017distance}. 
Since $Type\,SS$ eddies have heights $\mathcal{H} \sim \delta$, they do not contribute to the $w-$ spectra in the log region (figures \ref{wspectra_scaling}(d,h)).

While the current model, which is developed based on the scaling of the $u-$ spectra, captures the key scaling arguments of the $v-$ and the $w-$ spectra, we note that further modifications are required to better model the 2-D spectra of all components of velocity. For example, tuning the shape of the hairpins could possibly resolve the bimodal nature of the $v-$ spectra. However, such refinements would require 2-D spectra of $v$ and $w$ at high Reynolds numbers.

\section{Discussion on spectral self-similarity based on the extended AEM} \label{r3_disc}
The slope ($m$) of the 2-D spectra of $u$, which is equivalent to the ratio of the plateau in the 1-D streamwise spectra to the plateau in the 1-D spanwise spectra of $u$, has been reported by \citet{chandran2017JFM} to be an indicator of self-similarity. They observed the value of $m$ to monotonically increase with Reynolds number towards 1, with $m=1$ suggesting self-similarity. 
Here, using the extended AEM, we discuss a kinematic perspective on this empirically observed trend of $m$ with $Re_\tau$.

\begin{figure}
	\centering
	\includegraphics[trim = 0mm 0mm 0mm 0mm, clip, width=13.5cm]{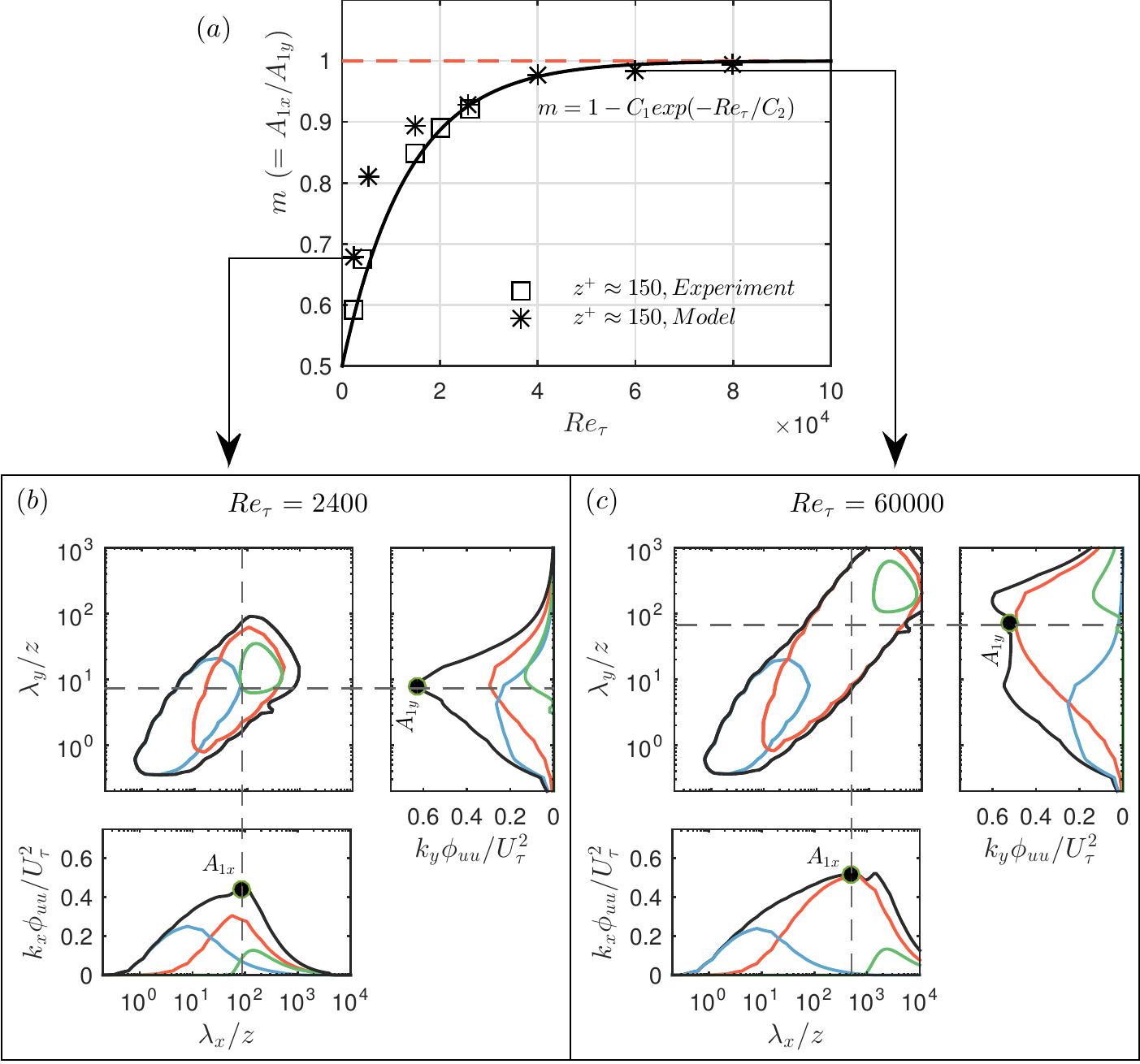}
	\caption{(a) Variation of $m$ versus $Re_\tau$ at $z^+\approx 150$ from experiments and the extended AEM. The solid black curve is the empirical fit of the form $m=1-C_1exp(-Re_\tau/C_2)$ from \citet{chandran2017JFM}, where $C_1=0.5$ and the value of $C_2$ is simply fitted to the data. (b) and (c) 2-D spectrum and the associated 1-D spectra at $Re_\tau=2400$ and $Re_\tau=60000$, respectively, from the extended AEM. The energy contribution of $Type\,C_A$ (blue), $Type\,A$ (red) and $Type\,SS$ (green) motions are plotted in (b) and (c).}
	\label{mvsRe}
\end{figure}

Figure \ref{mvsRe}(a) shows the plot of $m=A_{1x}/A_{1y}$ vs $Re_\tau$ at $z^+\approx150$ from both experiments and the extended AEM, and the red dashed line indicates self-similarity. The results from the extended AEM follow the empirically observed Reynolds number trend reasonably well. We note that the values are slightly over predicted at low Reynolds numbers while matching well with experiments for $Re_\tau \gtrsim 10^4$. Agreeing with the empirical fit, the value of $m$ obtained from the model is observed to approach unity at $Re_\tau \approx 60000$. Here, it is to be noted that a model comprising of $Type\,A$ eddies alone would always predict $m=1$ irrespective of Reynolds number (red dashed line). 

In order to understand the Reynolds number trend of $m$, we analyze the 2-D spectrum with the associated 1-D streamwise and spanwise spectra at $Re_\tau = 2400$ ($m\approx0.5$) and $Re_\tau = 60000$ ($m=1$), obtained from the extended AEM, shown in figures \ref{mvsRe}(b) and (c), respectively. The contributions of $Type\,A$, $Type\,C_A$ and $Type\,SS$ eddies are highlighted and color-coded in the figure. The plateaus in the streamwise and spanwise spectra, $A_{1x}$ and $A_{1y}$, are also highlighted.
At $Re_\tau=2400$ (figure \ref{mvsRe}b), there is less scale separation between the largest ($Type\,SS$) and the smallest ($Type\,C_A$) energetic motions which result in an overlap of the energy contributions from $Type\,C_A$, $Type\,A$ and $Type\,SS$ eddies for $\lambda_x>\mathcal{O}(10z)$ and $\lambda_y>\mathcal{O}(z)$. As discussed in \ref{AEMresults_spectra of u_text}, the $\lambda_y/z \sim (\lambda_x/z)^{1/2}$ relationship ($m=0.5$) at such length scales is observed to be a result of the overlap of sub-component energies. The 1-D streamwise and spanwise spectra are obtained by integrating the 2-D spectrum as given in equation \ref{eq_1dxspectra}. Therefore, the plateaus in the 1-D streamwise and spanwise spectra, $A_{1x}$ and $A_{1y}$ respectively, are obtained by integrating the 2-D spectrum along the vertical and the horizontal dashed lines in figure \ref{mvsRe}(b), respectively. At $Re_\tau=2400$, $A_{1x}$ and $A_{1y}$ have contributions from all three spectral subcomponents: $Type\,C_A$, $Type\,A$ and $Type\,SS$. Since $Type\,C_A$ energy diminishes beyond $[\lambda_x/z,\lambda_y/z]\sim 10$, its contribution to $A_{1x}$ at $\lambda_x/z\sim100$ is from its roll-off, and therefore is relatively low. However, for the plateau in the spanwise spectra $A_{1y}$, which is at $\lambda_y/z\sim10$, the contribution of $Type\,C_A$ is high and in proportion to that of $Type\,A$. Since $Type\,A$ and $Type\,SS$ contribute similarly to $A_{1x}$ and $A_{1y}$, the increased contribution from $Type\,C_A$ to $A_{1y}$ results in $A_{1y}>A_{1x}$ and therefore, $m<1$.

The scale separation between the largest and the smallest scales increases with Reynolds number. Referring back to figure \ref{Types_Scaling}, $Type\,C_A$ and the small-scale end of $Type\,A$ follow inner-flow scaling while $Type\,SS$ and the large-scale end of $Type\,A$ follow outer-scaling. Therefore, with increasing Reynolds number (or decreasing $z/\delta$), $Type\,SS$ spectra and the large-scale end of $Type\,A$ spectra shift to larger $\lambda_x/z$ and $\lambda_y/z$. As seen from figure \ref{mvsRe}(c), at $Re_\tau=60000$, $Type\,C_A$ and $Type\,SS$ spectra are completely separated from each other at the wavelengths corresponding to the locations of $A_{1x}$ and $A_{1y}$ which are $\lambda_x/z \approx 500$ and $\lambda_y/z \approx 70$ respectively. As a consequence, at this Reynolds number, $A_{1x}$ and $A_{1y}$ have energy contributions only from the wall-coherent self-similar $Type\,A$ motions (spectra in red). Hence, from figure \ref{mvsRe}(c), $\lambda_x/z \approx 500$ and $\lambda_y/z \approx 70$ represent the length scales at which a true $k^{-1}$ scaling commences in a 1-D streamwise and 1-D spanwise spectra, respectively. Even though a true $k^{-1}$ scaling kicks in at $Re_\tau \approx 60000$, a decade of $k^{-1}$ scaling may be revealed only at even higher Reynolds numbers.

\begin{figure}
	\centering
	\includegraphics[trim = 0mm 0mm 0mm 0mm, clip, width=12.5cm]{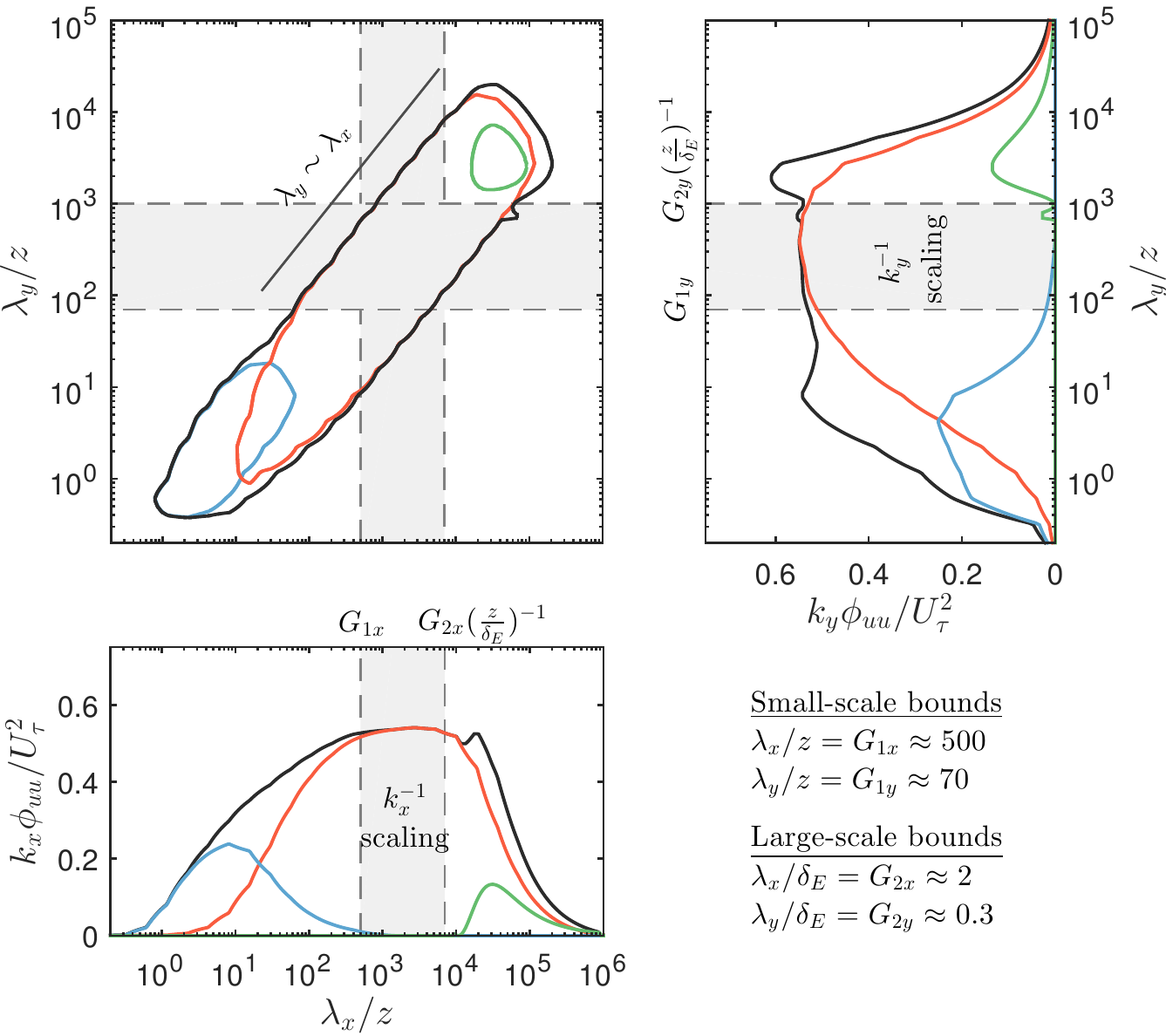}
	\caption{Spectra of $u$ at asymptotic high Reynolds number ($\mathcal{O}(10^6),\,z/\delta_E\sim10^{-4}$) highlighting a decade of $k^{-1}$ plateau in both streamwise and spanwise spectra. The small-scale and large-scale bounds of the $k^{-1}$ region is indicated in the plots. The energy contribution of  $Type\,C_A$ (blue), $Type\,A$ (red) and $Type\,SS$ (green) motions are highlighted.}
	\label{k-1_asymptote}
\end{figure}

The prediction from the model at an extreme $Re_\tau$ ($\mathcal{O}(10^6)$) is shown in figure \ref{k-1_asymptote}. A decade of $k^{-1}$ scaling is evident in both streamwise and spanwise spectra. The small-scale bound of the $k^{-1}$ region corresponds to the scales where the `large-scale roll-off' from $Type\,C_A$ energy ends. Since the roll-off scales with $z$, the $k^{-1}$ region begins at fixed inner-scaled wavelengths, $G_{1x}$ and $G_{1y}$ respectively in the 1-D streamwise and spanwise spectra. Based on the results from the extended AEM, these bounds are estimated to be  $\lambda_x/z=G_{1x} \approx 500$ and $\lambda_y/z=G_{1y} \approx 70$ respectively. The limit in the streamwise spectra agrees with \citet{baars2020a}, who estimated $G_{1x} \approx 385$. Similarly, the large-scale bound of the $k^{-1}$ region corresponds to the scales where the `small-scale roll-off' from $Type\,SS$ energy ends. Since the roll-off scales with $\delta_E$, the $k^{-1}$ region at the large-scales would be bounded by fixed outer-scaled wavelengths, $G_{2x}$ and $G_{2y}$ respectively in the 1-D streamwise and spanwise spectra. From figure \ref{k-1_asymptote}, these bounds are estimated to be  $\lambda_x/\delta_E=G_{2x} \approx 2$ and $\lambda_y/\delta_E=G_{2y} \approx 0.3$ respectively. Therefore, a decade of $k_x^{-1}$ scaling would require $G_{2x}(z/\delta_E)^{-1}/G_{1x}\sim10$, and a decade of $k_y^{-1}$ scaling would require $G_{2y}(z/\delta_E)^{-1}/G_{1y}\sim10$, or in both cases, $z/\delta_E\sim10^{-4}$ (figure \ref{k-1_asymptote}).

We note that the conclusions from the current model are based on a perfect outer-flow scaling of $Type\,SS$ energy. The work of \citet{baars2020a} reports very large scale energy contributions to have a subtle trend with Reynolds number. The authors, however, report the trend to be weaker than previous observations \citep{hutchins2007evidence,vallikivi2015spectral,vallikivi2015turbulent}. Even though the trend appears to be less significant within the log region $2.6Re_\tau^{1/2}\leq z^+\leq0.15Re_\tau$, the asymptotic predictions would benefit clarity on the outer-flow scaling arguments of very large scale motions.

\section{Summary and conclusions}
The attached eddy model (AEM) comprising only self-similar wall-attached eddies ($Type\,A$) is observed to represent the dominant large-scales in the logarithmic region only at high Reynolds numbers. 
However, when compared to experimental data, the energy left unresolved by these $Type\,A$ eddies is found to be significant enough to dictate the trends of the two-dimensional (2-D) spectra, even at Reynolds numbers as high as $Re_\tau=26000$.
Therefore, an extension to the AEM is proposed by incorporating into the model two additional types of structures that are major contributors to the turbulent kinetic energy in the logarithmic region: 
(i) $Type\,C_A$ eddies, representative of the wall-incoherent, small-scale structures that follow a self-similar distance from the wall scaling and 
(ii) $Type\,SS$ eddies, representative of the wall-coherent, very-large-scale (superstructure-like) motions or the global modes. The geometry of these representative eddies and their organization within the boundary layer are identified based on the experimentally observed inner-flow ($z$-scaling) and outer-flow scaling ($\delta$-scaling)  of the 2-D energy spectra of $u$. 

When considering the energy spectra of $u$, $v$ and $w$ that is obtained from the extended AEM, in addition to the energy contribution from $Type\,A$ eddies, there is the $z$-scaled energy contribution from $Type\,C_A$ eddies at $[\lambda_x \sim z,\lambda_y \sim z]$ and the $\delta$-scaled energy contribution from $Type\,SS$ eddies at $[\lambda_x \sim 10\,\delta,\lambda_y \sim \delta]$. Consequently,  the model captured the experimentally observed trends of the 2-D energy spectra of all three velocity components reasonably well, across a greater range of energetic scales from $\mathcal{O}(z)$ to $\mathcal{O}(10\,\delta)$. The model also captured the empirically observed shift in the trend of the energetic large-scales in the 2-D spectra of $u$, from a $\lambda_y/z \sim (\lambda_x/z)^{1/2}$ relationship at low-Reynolds numbers towards the self-similar $\lambda_y \sim \lambda_x$ scaling at high-Reynolds numbers. 
A discussion on this Reynolds number trend is presented for the spectra of $u$, based on which, self-similarity would be evident with a $\lambda_y \sim \lambda_x$ scaling for a region of constant energy in the 2-D spectrum, and the associated $k^{-1}$ scaling in the 1-D streamwise and spanwise spectrum, only at $Re_\tau \gtrsim 60000$, when a complete scale-separation between the $\delta$-scaled $Type\,SS$ and the $z$-scaled $Type\,C_A$ eddies is predicted to exist.

\begin{acknowledgements}
The authors gratefully acknowledge the financial support of this research from the Australian Research Council. 
The authors wish to thank Myoungkyu Lee and Robert Moser for providing the 2-D spectra from turbulent channel DNS at $Re_\tau = 5200$, and Woutijn Baars, Rio Baidya, Charitha de Silva and Rahul Deshpande for helpful discussions and feedback on the manuscript.
\end{acknowledgements}

\bibliography{mybibfile2}

\end{document}